\documentclass[a4paper,12pt]{article}
\usepackage{soul}
\usepackage[usenames,dvipsnames]{color}
\usepackage{jheppub}
\usepackage{esvect}
\usepackage{amsmath, amssymb, slashed, epsf, color, graphicx, latexsym, tensor, physics, xcolor, array, tabularx, makecell,bm, tikz, enumitem, longtable, float}

\usepackage{epsfig}
\usepackage{comment}
\usepackage{graphics}
\usepackage{mathtools}
\usepackage{appendix}
\usepackage{inputenc}
\usepackage{indentfirst}

\newcommand{\ub}[1]{\underline{\bm{#1}}}
\usetikzlibrary{decorations.pathmorphing, arrows.meta, calc}
\usetikzlibrary{decorations.pathmorphing, positioning}

\title{\scalebox{0.97}[1]{Quantum Love and ringdown near extremality}}
\author[1,2]{Arindam Bhattacharjee}
\author[3]{, Muktajyoti Saha}
\author[3]{, M. V. S. Saketh}

\affiliation[1]{Instituto de Ciencias Exactas y Naturales (ICEN), Universidad Arturo Prat,
Playa Brava 3256, 1111346 Iquique, Chile}
\affiliation[2]{Facultad de Ciencias, Universidad Arturo Prat,
Avenida Arturo Prat Chacón 2120, 1110939 Iquique, Chile}
\affiliation[3]{International Centre for Theoretical Sciences, Tata Institute of Fundamental Research,
Bengaluru 560089, India}

\emailAdd{arindamb.hep@gmail.com}
\emailAdd{muktajyoti.hepth@gmail.com}
\emailAdd{venkata.saketh@icts.res.in}

\abstract{In this paper we discuss the quasinormal modes and the tidal responses of a four-dimensional asymptotically flat Reissner-Nordstr\"{o}m black hole near extremality and investigate the nontrivial near-horizon quantum corrections of these observables. The near-horizon dynamics of near extremal black holes are generally governed by an effective Schwarzian action coupled to boundary gauge modes. We use this description and couple this effective theory with a massless scalar field outside. Using this setup we compute $2 \to 2$ scattering process of a black hole and a scalar particle. The poles of the scattering amplitude at complex frequencies give us the quasinormal mode spectrum and we recover the predicted zero damped modes near extremality at leading order. Next we evaluate the static Love numbers from the elastic limit of the scattering amplitude. We show that the static Love numbers are non-zero in contrast to the classical expectations. Very close to extremality, the behaviour is analytic, but as the near-extremal energy becomes sufficiently large, we find logarithmic dependence on the energy. This cannot be removed by the choice of renormalisation scheme. Thus near-horizon quantum effects non-trivially change the tidal response of near-extremal black holes.}

\begin{document}
\maketitle

\section{Introduction}
In recent years, Black Hole spectroscopy has become a viable way of testing predictions of General Relativity (GR) and its possible deviations \cite{Berti:2025hly}. Current experiments like LIGO and Virgo collaborations routinely observe black hole mergers and the post merger ringdown of the resulting black holes \cite{LIGOScientific:2016aoc}. The increasing sensitivity of these and future experiments gives us hope to test deviations from GR predictions in the black hole backgrounds. Oscillating black holes, like the ones created in a merger, radiate gravitational and electromagnetic waves on their way to become a stationary solution. Since black holes are by nature dissipative, the linearized perturbations of gravitational field or any other field in the black hole spacetimes have complex frequency quasinormal modes (QNM) \cite{Berti:2009kk}. These QNMs are characteristic properties that carry the information about the mass, charge and the spin of the black hole. In general, solving the spectrum of the quasinormal modes is an extremely difficult problem \cite{Leaver:1985ax, Detweiler:1980gk}. However some simplifications may arise on certain special sectors of the gravitational phase space. One such sector is the phase space near extremality where the surface gravity of the black holes are very small \cite{Hod:2008zz, Hod:2008se}.  

While QNMs describe how a black hole rings down after a merger, the inspiral phase
probes a complementary property: the response of the black hole to the static tidal
field of its companion. This response is quantified by the tidal Love numbers, which
relate the multipole moments induced in the body to the external tidal field. For neutron stars they encode
the equation of state, but for black holes in four-dimensional General Relativity the
static Love numbers vanish identically, as shown first for Schwarzschild
\cite{Binnington:2009bb,Damour:2009vw, Hui:2020xxx, Parra-Martinez:2025bcu} and later for Kerr black holes
\cite{Charalambous:2021mea, Charalambous:2021kcz}. A useful framework to understand tidal deformations of black holes is the Worldline Effective Field Theory (WEFT), where the black hole is regarded as a point particle described by a local worldline action. In this framework, Love numbers are defined as Wilson coefficients
of finite-size operators on the worldline of the black hole \cite{Goldberger:2004jt,Kol:2011vg}. Beyond the static limit, black holes exhibit a nonvanishing tidal response, described
by dynamical (frequency-dependent) Love numbers and by dissipative effects associated
with absorption at the horizon \cite{Goldberger:2005cd,Saketh:2023bul}. For a general review of Love numbers see \cite{Rodriguez:2026iot, Chakraborty:2026qru}. Since the vanishing of black hole static Love numbers is an artifact of emergent symmetries in GR, it does not generally hold in modified
theories of gravity or for horizonless compact objects \cite{Horowitz:2023xyl, Cardoso:2017cfl, Cardoso:2018ptl}. Quantum corrections can also cause non-vanishing static Love numbers \cite{Kim:2020dif, Elkhidir:2026lgp}. Hence, non-vanishing static
Love numbers constitute a clear signature of new physics, a sharp test of GR in the strong-field regime. The main goal of this paper is to compute quantum gravitational corrections to static Love numbers and QNMs of black holes very close to extremality. 

Some of the black holes in our universe are expected to be rapidly rotating i.e., \emph{very close to maximally rotating Kerr black holes} (e.g. M87* at the centre of the constellation Virgo has been inferred to have a high spin) and share key qualitative features with \emph{maximally charged Reissner-Nordstr\"{o}m black holes}. Extremal black holes are black holes that have maximum charge or spin given their mass. They have zero surface gravity and hence zero Hawking temperature. Near-extremal black holes are solutions that slightly deviate from the extremality conditions. Typically when the energy of the black hole above extremality is of the order $E_0 = G_N/r_0^3$, where $r_0$ is the radius of the extremal black hole, the near-extremal black holes behave significantly different from typical non-extremal black holes. In this limit, the near-horizon geometry universally factorises into a large two-dimensional throat region, where the metric is nearly $AdS_2$. A feature of this universality is that the dynamics of near-extremal black holes are significantly controlled by fluctuations at the boundary of this throat. At strict extremality, the thermodynamics of the black holes are governed by fluctuations that follow asymptotic $AdS_2$ boundary conditions \cite{Sen:2007qy, Sen:2008vm, Sen:2008yk}. In presence of a small temperature, the thermodynamics gets dominant contributions from special type of modes, which are pseudo Goldstone modes associated with the asymptotic symmetries of the $AdS_2$ geometry. A universal sector of these contributions arise from large diffeomorphisms or time reparametrisations of the $AdS_2$ boundary. Interestingly, the soft breaking of these asymptotic symmetries can be effectively described by one-dimensional modes localised at the boundary of $AdS_2$ \cite{Nayak:2018qej, Iliesiu:2020qvm}. This effective theory is described in terms of a Schwarzian sector corresponding to the breaking of large diffeomorphisms of the throat and some coupled gauge fields corresponding to other isometries and gauge fields in the four dimensional spacetime as well. In recent years, this effective description of near-extremal black holes has shed light in many theoretical aspects of these black holes. For instance, they significantly affect the entropy of near-extremal black holes by introducing logarithmic temperature corrections \cite{Iliesiu:2020qvm, Iliesiu:2022onk, Banerjee:2023quv, Banerjee:2023gll, Kapec:2023ruw, Maulik:2024dwq, Maulik:2025phe, Blacker:2025zca}. They also modify the thermal nature of the Hawking radiation thereby changing the evolution of these black hole systems \cite{Brown:2024ajk, Maulik:2025hax, Bhattacharjee:2025wfv} and modify absorption amplitudes \cite{Emparan:2025sao, Emparan:2025qqf, Luo:2026epp}, to name a few\footnote{Recently, their uplift to the full black hole geometry has been discussed in \cite{Kolanowski:2024zrq, Castro:2025itb}}. We will use this effective description of quantum gravitational fluctuations to compute corrections to static Love numbers and QNMs.



The observationally relevant black holes are spinning Kerr black holes that emit strong gravitational waves during its ringdown phase. However, this is a very difficult problem due to the lack of spherical symmetry of the background. In this work, we simplify the problem by considering charged Reissner-Nordstr\"{o}m (RN) black holes that have spherical symmetry and since they address the quantum corrections that we are interested in. We also consider a massless scalar field in the bulk to simplify the problem. However the qualitative features of the results we find will still be valid in the case of gravitational radiations emitted by near-extremal Kerr black holes. In this paper we will model the near-extremal RN black hole as a Schwarzian system coupled to a $SU(2)$ gauge mode placed on a timelike geodesic in flat background. This system couples to an external massless scalar field via non-normalisable modes sourced by the external field that are localised at the boundary of the $AdS_2$ throat. The asymptotic states of the full system at $t \xrightarrow{} \pm \infty$ is assumed to be tensor products of the Schwarzian+$SU(2)$ states with the free particle states of the external scalar field. The $SU(2)$ gauge fields are necessary for incorporating the interactions between the BH and scalar quanta that carry nonzero angular momenta. In particular it gives rise to selection rules between different angular momentum sectors.

Our primary objective would be to study a $2 \xrightarrow{}2$ scattering amplitude between the BH and a scalar quantum. However, we need to be very careful since the first thing that this Schwarzian picture predicts is that the black hole system is unstable and naturally decays via emitting scalar particles. This decay is nothing but the Hawking radiation but the nature of the decay near extremality differs significantly from the usual semiclassical thermal emission \cite{Brown:2024ajk}. We calculate the leading order behaviour of this decay rate and use it to modify our propagators in the scattering process. We find that the decay is very slow.

Firstly, we compute the quasinormal modes of the system by analysing the poles of the integrand of the scattering amplitude and studying their overlaps by analytically continuing to complex frequencies. We recover the known results of zero damped quasinormal modes \cite{Zimmerman:2015trm, Hod:2008se, Mukherjee:2024nhx} for the near-extremal black holes in the semiclassical regime. Next, we find the QNMs for black holes in the quantum regime as predicted by the effective description. We find that in leading order the quasinormal modes of this system stay the same but their functional form in terms of temperature changes.

Finally, we compute the Love number by studying the \textit{conservative response} of the amplitude when the energy of the black hole remains unchanged. In this regime, we compute the real part of the amplitude and from it extract the static Love numbers. We see that for both classical and quantum black holes, the $l = 0$ Love numbers vanish. This is consistent with the shift symmetry of the massless scalar action. For non-zero $l$ modes, we find non-trivial response in both semiclassical and quantum black hole backgrounds with generally divergent pieces. We regulate these divergences by adding appropriate counterterms and show that there are nontrivial tidal responses after a regularisation procedure has been chosen.

The rest of the paper is organised as follows: In section \ref{sec: Sch-review}, we review the basic aspects of the effective description of the near-horizon dynamics of near-extremal black holes and lay out the rules and conventions to be followed in the paper. In section \ref{sec: decay}, we compute the leading decay rate of a black hole in semiclassical and quantum regimes. In section \ref{sec: scattering}, we find the general form of the $2\to 2$ scattering amplitude. Using this, we evaluate the semiclassical and quantum QNMs in section \ref{sec: qnm}. Finally, we compute the non-vanishing static Love numbers in semiclassical and quantum regimes in section \ref{sec: Love}. These two sections comprise of the most important results of this paper. We have two appendices \ref{App: identities} and \ref{App: Hamiltoninan} with supporting computations.

\section{Effective model: Schwarzian coupled to free QFT}\label{sec: Sch-review}

In this section, we will briefly review some aspects of the effective description of near-extremal black holes and set up the conventions to be used in our computations. The low energy near-horizon dynamics of near-extremal black holes can be effectively governed by a one-dimensional theory living on the boundary of the near-horizon region. This theory emerges as the effect of explicit and slight breaking of asymptotic symmetries of $AdS_2$. A universal sector of this theory is the Schwarzian action appearing from the breaking of time reparametrizations of the boundary \cite{Iliesiu:2020qvm}. In this paper, we are interested in a charged near-extremal black hole that can be described by two large scales: the extremal horizon size or charge $q = r_0$ and the inverse temperature $\beta$. These parameters satisfy the condition $\beta \gg r_0$, which signifies that the separation between the two horizons of the black hole is much smaller compared to the horizon scale $r_+ - r_- \ll r_0$. The mass of the black hole is given as $G_N M = r_0 + \mathcal{O}(E_0^{-1}T^2)$. The effective theory is governed by a coupling, $$E_0 = \frac{G_N}{r_0^3},$$where $G_N$ is the 4D Newton constant. This energy scale $E_0$ signifies the breakdown of semiclassical physics. The effective theory has an upper and a lower regimes of validity. At the high energies, it breaks down when $\beta\sim r_0$ signaling a violation of the near-extremality condition. At exponentially low energies, for instance, when $\beta \sim r_0 e^{S_0}$, with $S_0 \sim r_0^2/G_N$ being the extremal entropy, the theory necessarily breaks down. Naturally, the analyses of this paper are valid within these energy scales.

Our major computations will be done in the microcanonical ensemble, where we will parametrize the black hole with its available energy above the ground state $E$, dual to $\beta$. The ground state energy for a given angular momentum sector $L$ is $r_0 + E^0_L$, which we define below. The quantum gravitational corrections are suppressed when $E\gg E_0$ (while staying near extremality), however, they become dominant as $E\lesssim E_0$. These limits are called the \textit{semiclassical} and \textit{quantum} regimes respectively. In our case, the boundary theory also has additional modes coming from the large $SU(2)$ symmetries associated with the $S^2$ factor of the geometry\footnote{We will couple an uncharged field with the boundary theory and thus the large $U(1)$ symmetries associated with the 4D Maxwell field will not play a role.}. Thus, the near-horizon dynamics is described by the Hamiltonian \cite{Heydeman:2020hhw},
\begin{align}\label{free-Sch}
    H_{0} \equiv H_{\rm Sch} + H_{SU(2)} = E_0^{-1}
\left[\operatorname{Sch}(f,t) + \operatorname{Tr}\left(g^{-1}\partial_t g\right)^2\right],
\end{align}
Here $f(t)$ is the time reparametrization mode and ${\rm Sch} (f,t) = \frac{f'''}{f'}
-\frac{3}{2}\left(\frac{f''}{f'}\right)^2$ is its Schwarzian derivative. $g(t) \in SU(2)$ is the $SU(2)$ mode governed by a particle on a group manifold action. The states in this theory are given as,
\begin{align}
    H_{0} \ket{E,L,M} = (E + E^0_L) \ket{E,L,M}, \quad E^0_L \equiv \frac{L(L+1)}{2}E_0.
\end{align}
Here, $E$ is the available energy of the state in the $L$-angular momentum sector, while $E^0_{L}$ appears from the $SU(2)$ part of the Hamiltonian. In the semiclassical limit, this threshold energy of a particular angular momentum sector is not visible, while it becomes important in the quantum regime. The density of states around fixed $(E,L,M)\equiv\ub{E}$ is given as\footnote{Note that, the factor of $(2L+1)$ has appeared due to the dimension of the spin-$L$ representation, and not due to a sum over azimuthal quantum number.},
\begin{align}
    \rho(\ub{E}) = E_0^{-1}(2L+1) e^{S_0} \rho_S(E) = \frac{e^{S_0}}{2\pi^2 E_0}(2L+1) \sinh\left(2\pi\sqrt{\frac{2E}{E_0}}\right)\Theta(E).
\end{align}
The states follow the normalisation and completeness condition,
\begin{align}
    &\int dE \sum_{LM} \rho(\ub{E})\ket{E,L,M}\bra{E,L,M} = 1, \\
    &\bra{E,L,M}\ket{E,L,M} = \rho(\ub{E})^{-1}\delta(E-E')\delta_{LL'}\delta_{MM'}.
\end{align}

Our goal is to study the scattering of an uncharged spin-0 particle on the near-extremal black hole background and incorporate the effects of the quantum gravitational modes stemming from its near-horizon geometry. For this, we couple the effective theory \eqref{free-Sch} with an external scalar field, following \cite{Brown:2024ajk}. The main idea is to  separate the near-extremal geometry into two large near-horizon and far-horizon regions and to solve the scalar field equation in these regions. In the far region, the scalar field is well approximated to a free field in flat space.    Owing to the time translation and $S^2$ isometry of the background, the problem can be recast into solving a radial differential equation in the two regions. The coupling between the Schwarzian+$SU(2)$ system and the far-horizon free QFT can be derived by noting that the free QFT has a particular mode that behaves as a non-normalizable or source corresponding to $AdS_2$. This mode sources an operator of particular representation in the Schwarzian+$SU(2)$ theory following $AdS_2/CFT_1$. Thus, it induces an operator-source interaction in the boundary theory. We present the details of this computation in the appendix \ref{App: Hamiltoninan}. The effective Hamiltonian of the system at the near-horizon boundary is given by,
\begin{align}\label{int-hamil}
    & H = H_{0} + H_{I}(t), \\
    & H_{I}(t) =  \sum_{lm}\mathcal{N}_l e^{i H_{0}t} \mathcal{O}_{\Delta lm} e^{-i H_{0}t}\int_{0}^{\infty} d\omega \phi_{\omega lm} (t); \quad \mathcal{N}_l = r_0^{2l+1}E_0^{\Delta}.
\end{align}
Here $\mathcal{O}_{\Delta lm}$ is an operator in the boundary theory with conformal dimension $\Delta = l+1$ and $SU(2)$ labels $l,m$. Whereas, $\phi_{\omega lm}$ is proportional to the mode of the QFT that falls off as $r^l$ near the $AdS_2$ boundary and it is given by,
\begin{align}\label{source}
    & \phi_{\omega lm}(t) = u_{\omega lm} e^{-i\omega t}a_{\omega lm} + u^*_{\omega lm} e^{i\omega t}a^{\dagger}_{\omega lm} \equiv \phi_{\omega lm}^{+} + \phi_{\omega lm}^{-}, \\
   & u_{\omega lm} = \frac{\omega^{l+1/2}}{\Gamma(l+3/2)}(2i)^{-l-1}.
\end{align}
The factors of $r_0$ in the normalization $\mathcal{N}_l$ has appeared from near-far matching conditions of the scalar field, while the factors of $E_0$ are just extracted from the operator $\mathcal{O}_{\Delta lm}$, making it dimensionless. $a_{\omega lm}, a^{\dagger}_{\omega lm}$ are the asymptotic modes of the QFT. They enter the Hamiltonian through classically matching the solutions at the asymptotic and intermediate regions. They satisfy,
\begin{align}
    \left[a_{\ub{\omega}}, a^{\dagger}_{\ub{\omega}'}\right] = \hat{\delta}(\ub{\omega},\ub{\omega}') = \delta(\omega-\omega')\delta_{ll'}\delta_{mm'}.
\end{align}
For convenience, sometimes, we will use the boldface underline to indicate the three representation labels of energy, azimuthal and magnetic quantum numbers of a particular object.

\subsection{Feynman rules}\label{rules}
In our computations, we will need to compute correlation functions in this effective system with appropriate operator insertions. For this, let us briefly write down the relevant Feynman rules for the Schwarzian theory with $SU(2)$ matter coupling following \cite{Mertens:2017mtv, Blommaert:2018oro}. The rules apply to the insertion of bilocal operators. One end of such a bilocal operator i.e. each vertex of incoming state $\ub{E}_1$ and outgoing state $\ub{E}_2$ with an operator $\ub{\Delta}$ correspond to a factor of, \\

\begin{minipage}{0.5\textwidth}
\centering
\begin{tikzpicture}[thick]

\draw (-2,0) arc (180:0:2);

\fill (0,2) circle (2pt);

\draw (0,2) -- (0,0.1);

\node[below] at (-2,0) { $\ub{E}_1$ };
\node[below] at (2,0) { $\ub{E}_2$ };

\node[right] at (0,1.1) {$\ub{\Delta}$};


\end{tikzpicture}
\end{minipage}
\hfill
\begin{minipage}{0.5\textwidth}
\centering
\begin{align*}
    = \gamma_{L_1M_1,L_2M_2; lm}\, \left(\frac{\Gamma^{\Delta}_{12}}{2^{2\Delta-1}e^{S_0}\Gamma(2\Delta)}\right)^{1/2}.
\end{align*}    
\end{minipage}
Here the contribution from the Schwarzian sector is given by,
\begin{align}\label{Sch-vertex}
    \Gamma^{\Delta}_{12} = \prod_{\pm}\Gamma(\Delta \pm i \sqrt{2E_0^{-1}E_1}\pm i \sqrt{2E_0^{-1}E_2}).
\end{align} 
Another form of this factor is given in \eqref{Sch-vertex-simple}. The SU(2) sector gives,
\begin{align}\label{su(2)-vertex}
\gamma_{L_1M_1,L_2M_2; lm} =
    \begin{pmatrix}
    L_1  & l & L_2\\
    M_1  & m & M_2
\end{pmatrix},
\end{align}
which is the Wigner 3j-symbol. This factor comes because three representations are mixing at the vertex. While writing the columns of the 3j-symbol, we have considered a clockwise orientation across the vertex. When we consider the diagrams with bilocal operators, then at the other end of the operator, the orientation should be anti-clockwise. This also naturally incorporates selection rules at the vertex,
\begin{align}\label{selection-rule}
    \abs{L_1 - L_2} \leq l \leq L_1 + L_2, \quad M_1 + M_2 + m = 0.
\end{align}
If these rules are not satisfied, the symbol gives zero, ruling out such processes.

In higher point diagrams whenever there is a crossing of two lines within the disc, it introduces an extra factor called the crossing kernel:\\

\begin{minipage}{0.48\textwidth}
\begin{center}
\begin{tikzpicture}[scale=0.6]


\draw[thick,dotted] (2,0) circle (2);

\node[above] at (2,2.2) {$\ub{E}_2$};
\node[below] at (2,-2.2) {$\ub{E}_1$};

\node[left]  at (-0.4,0) {$\ub{E}'$};
\node[right] at (4.4,0) {$\ub{E}$};

\coordinate (TL) at (0.85, 1.63);
\coordinate (TR) at (3.15, 1.63);
\coordinate (BL) at (0.85,-1.63);
\coordinate (BR) at (3.15,-1.63);

\foreach \p in {TL,TR,BL,BR}{
    \draw[thick,dashed] ($(\p)+(-0.10,-0.10)$) -- ($(\p)+(0.10,0.10)$);
    \draw[thick,dashed] ($(\p)+(-0.10,0.10)$) -- ($(\p)+(0.10,-0.10)$);
}

\node[left]  at ($(TL)+(-0.25,0.05)$) {$\mathcal{O}_{\ub{\Delta}_1}$};
\node[right] at ($(TR)+(0.25,0.05)$) {$\mathcal{O}_{\ub{\Delta}_2}$};

\node[left]  at ($(BL)+(-0.25,-0.45)$) {$\mathcal{O}_{\ub{\Delta}_2}$};
\node[right] at ($(BR)+(0.25,-0.45)$) {$\mathcal{O}_{\ub{\Delta}_1}$};

\draw[thick]
(TL) -- (BR);

\draw[thick]
(BL) -- (TR);

\end{tikzpicture}
\end{center}
\end{minipage}
\hfill
\begin{minipage}{0.48\textwidth}
\begin{align}\label{cross}
= \begin{Bmatrix}
    \Delta_2 & E_2 & E \\
    \Delta_1 & E_1 & E'
\end{Bmatrix} 
\begin{Bmatrix}
    l_2 & L_2 & L \\
    l_1 & L_1 & L'
\end{Bmatrix}.
\end{align}

\end{minipage}
The dashed circle means, we are only considering the contribution from the crossing. After this crossing kernel is taken into account, the rest of the diagram is to be thought of as an uncrossed one and for each vertex, we have to insert the appropriate vertex factor.

The crossing kernel consists of 6j-symbols associated with the Schwarzian part and the SU(2) representations. Their explicit form is given by,
\begin{align}\label{Sch-6j}
\begin{Bmatrix}
\Delta_1 & E_2 & E_1 \\
\Delta_2 & E_4 & E_3
\end{Bmatrix}
=&
\Big[\Gamma^{\Delta_1}_{12}\Gamma^{\Delta_2}_{23}\Gamma^{\Delta_1}_{34}\Gamma^{\Delta_2}_{41}
\Big]^{\frac12}\Gamma(2\Delta_1)\Gamma(2\Delta_2)
\nonumber\\
\times
&\mathbb{W}\!\left(
k_1,
k_3;
\Delta_1+ik_4,
\Delta_1-ik_4,
\Delta_2-ik_2,
\Delta_2+ik_2
\right).
\end{align}
Here, $\mathbb{W}$ is the Wilson function and $k_i \equiv \sqrt{2E_0^{-1}E_i}$ are the available momenta associated with the states. It is defined as,
\begin{align}\label{wilson}
    \mathbb{W}(\alpha,\beta;a,b,c,d)
\equiv
\frac{
\Gamma(d-a)\,
{}_4F_3\!\left(
\begin{matrix}
a+i\beta,\; a-i\beta,\; \tilde{a}+i\alpha,\; \tilde{a}-i\alpha \\
a+b,\; a+c,\; 1+a-d
\end{matrix}
;1
\right)
}{
\Gamma(a+b)\Gamma(a+c)\Gamma(d\pm i\beta)\Gamma(\tilde{d}\pm i\alpha)
}
+ (a \leftrightarrow d),
\end{align}
here $2\tilde{d} = b+c+d-a, 2\tilde{a} = b+c+a-d$. We discuss various properties of this function in the appendix \ref{App: identities}. By construction, the vertex factors and the crossing kernel are real.

These rules will allow us to compute the relevant scattering amplitude in the effective theory, but first, we will focus on the decay rate of a state which will be important for the amplitude.

\section{Decay rate}\label{sec: decay}

In this section, we will compute the decay rate of a black hole with available energy $E$ and angular momentum $L$ which is represented by a state $\ket{E,L,M}$ in the Schwarzian+$SU(2)$ theory. In the free theory, the time evolution of the state is trivial, however, due to the weak coupling with the external QFT, the state slowly decays. The decay can be computed using first-order perturbation theory induced by the interaction \eqref{int-hamil}. Let us consider a process where initially at $t=-\infty$, the state of the combined system is $\ket{\ub{E}}\times\ket{0}$ and it goes to a final state $\ket{\ub{E}_f}\times\ket{\ub{\omega}}$ at $t=+\infty$. The amplitude of this process is given by,
\begin{align}
    \mathcal{A}(\ub{E}\to\ub{E}_f,\ub{\omega}) = -i\int_{-\infty}^{\infty} dt \sum_{l'm'}\mathcal{N}_{l'}&\int d\omega' \mathcal{O}_{\ub{\Delta}'}(\ub{E}_f,\ub{E}) e^{it\left(E_f + E^0_{L_f} - E - E^0_L\right)} \nonumber \\
    &\bra{\ub{\omega}}(\phi^{+}_{\ub{\omega}'}(t) + \phi^{-}_{\ub{\omega}'}(t))\ket{0}.
\end{align}
The exponential factors have appeared from the eigenvalues of the free time evolution operator. Again, according to our conventions, we have denoted: 
\begin{align*}
    & \ub{E}\equiv (ELM), \ub{E}_f \equiv (E_f L_f M_f), \ub{\omega} \equiv (\omega l m), \ub{\omega}' \equiv (\omega' l' m'), \\
    &\ub{\Delta} \equiv (l+1,l,m), \ub{\Delta}' \equiv (l'+1,l',m').
\end{align*}
Here, $\mathcal{O}_{\ub{\Delta}'}(\ub{E}_f,\ub{E}) = \bra{\ub{E}_f}\mathcal{O}_{\ub{\Delta}'}\ket{\ub{E}}$ is the matrix element of the operator and $\ket{\ub{\omega}}$ is a one-particle state in the QFT. In principle, we should have included all possible final states in the QFT, but since the interaction Hamiltonian is linear in creation/annihilation operators, only one-particle states of the QFT can contribute to the process at first-order perturbation theory. The flat space part of the amplitude simplifies to,
\begin{align}
    u^{*}_{\ub{\omega}}e^{i\omega t}\delta_{ll'}\delta_{mm'}.
\end{align}
These Kronecker delta constraints set $(l'm') = (lm)$ and collapse the angular momentum sum. We also get $\ub{\Delta}' = \ub{\Delta}$ such that the amplitude becomes,
\begin{align}
    \mathcal{A}(\ub{E}\to\ub{E}_f,\ub{\omega}) = -2\pi i\mathcal{N}_l u^{*}_{\ub{\omega}}\mathcal{O}_{\ub{\Delta}}(\ub{E}_f,\ub{E})  \delta\left(E_f + E^0_{L_f} + \omega - E - E^0_L \right).
\end{align}
Taking the squared amplitude and dividing by the infinite time integral, we get the decay rate,
\begin{align}
    \gamma(\ub{E}) = 2\pi\int d\ub{\omega} \int d\ub{E}_f \rho(\ub{E}_f)\mathcal{N}^2_l\abs{u_{\ub{\omega}}\mathcal{O}_{\ub{\Delta}}(\ub{E}_f,\ub{E}) }^2\delta\left(E_f + E^0_{L_f} + \omega - E - E^0_L \right).
\end{align}
At the last step, we have also summed over the final states since we are only interested in the decay of a given initial state. We have also used the notation: $\int d\ub{\omega} \equiv \int d\omega \sum_{lm}$ and $\int d\ub{E}_f \equiv dE_f \sum_{L_fM_f}$. Due to the delta function constraint, the $E_f$ integral projects to,
\begin{align}
    E_f = E-\omega + \left(\frac{L(L+1)}{2}
    -\frac{L_f(L_f+1)}{2}\right)E_0 \geq 0.
\end{align}
The positivity of the available energy puts constraints on the final state that have to be satisfied for the process to make sense \cite{Emparan:2025qqf}. The Schwarzian part of the decay rate is now given by the diagram,\\

\begin{minipage}{0.3\textwidth}
    \begin{tikzpicture}[scale=0.6]

\draw[thick] (0,0) circle (2);

\draw[thick,dashed] (-2,0) -- (2,0);

\draw[line width=1.5pt] (-2.14,-0.22) -- (-1.86,0.22);
\draw[line width=1.5pt] (-2.14,0.22) -- (-1.86,-0.22);

\draw[line width=1.5pt] (1.86,-0.22) -- (2.14,0.22);
\draw[line width=1.5pt] (1.86,0.22) -- (2.14,-0.22);

\node[above] at (0,2.2) {$\ub{E}_f$};
\node[below] at (0,-2.2) {$\ub{E}$};

\node[left] at (-2.45,0) {$\mathcal{O}_{\ub{\Delta}}$};

\node[right] at (2.45,0) {$\mathcal{O}_{\ub{\Delta}}$};

\end{tikzpicture}
\end{minipage}
\hfill
\begin{minipage}{0.7\textwidth}
    \begin{align}
        =\frac{\Gamma^{\Delta}_{E_f E}}{2^{2\Delta-1}e^{S_0}\Gamma(2\Delta)} 
        \begin{pmatrix}
    L_f & L & l \\
    M_f & M & m
\end{pmatrix}^2.
    \end{align}
    
\end{minipage}
We have used the Feynman rules \eqref{Sch-vertex} and \eqref{su(2)-vertex}\footnote{We have used two opposite orientations at the two vertices of the bilocal operator as mentioned in section \ref{sec: Sch-review}. This convention is consistently followed in the higher point diagrams as well.}. We perform the $M_f,m$ sums using the property \eqref{3j-M-sum} of the Wigner 3j-symbol. This gives us,
\begin{align}\label{decay-rate}
    \gamma(E,L) = 2\pi\int_0^\infty d\omega  \sum_{L_f,l} \frac{2L_f+1}{2L+1}\frac{\mathcal{N}_l^2 E_0^{-1}\omega^{2l+1}}{\pi^24^{2\Delta}\Gamma(2\Delta)(\Gamma(l+3/2))^2}\rho_S(E_f)\Gamma^\Delta_{E_f E},
\end{align}
with $E_f = E-\omega +E^0_L - E^0_{L_f}\geq 0$. Note that, $L_f,L,l$ satisfy the angular momentum selection rule and $\Delta = l+1$. This rate is independent of the magnetic quantum number $M$ of the state $\ub{E}$. Now, we will analyse the decay rate in semiclassical and quantum regimes.

\subsection{Semiclassical decay rate}
In the semiclassical limit, we assume that the available energies of all the Schwarzian states are much larger compared to the coupling $E_0$. Thus, we find that the angular momentum-dependent thresholds can be neglected and the spectra are continuously connected to extremality. Thus, in \eqref{decay-rate}, we can approximate,
\begin{align}
    E_f \approx E_f = E - \omega.
\end{align}
In this limit, the two-point function can be simplified using \eqref{Sch-vertex-simple},

\begin{align}
    \Gamma^\Delta_{E_f E}\approx\, &2\pi^2 4^{l} (2E_0^{-1}E)^{l}(l!)^2 \prod_{j=1}^l \left(1+\frac{\beta_E^2\omega^2}{4\pi^2j^2}\right) \nonumber \\
    &\frac{2\cdot 2E_0^{-1}(-\omega)}{\exp\left(2\pi\sqrt{2E_0^{-1}(E-\omega)}\right) - \exp\left(2\pi\sqrt{2E_0^{-1}E}\right)}.
\end{align}

Also, since the energy is semiclassical, the Schwarzian density of states takes the typical exponential form,
\begin{align}
    \rho_S(E-\omega) \approx \frac{1}{2}\exp\left(2\pi\sqrt{2E_0^{-1}(E-\omega)}\right).
\end{align}
This has nonzero support for all energies and the exponent is just the classical entropy above extremality corresponding to the final state\footnote{Technically, the Schwarzian effecive description is valid only in the near-extremal regime which puts a UV cutoff in the energy scale. However, in this case the integrand dies off sufficiently fast at high energies and hence the cutoff does not play any role here. We will see in the next section that the cutoff enters in a particular scattering computation.}. We have also used $\omega\ll E$ limit. Now, the decay rate \eqref{decay-rate} simplifies to,

\begin{align}
    \gamma_{\rm sc}(E,L) \simeq 2\pi &\sum_{L_f,l} \frac{2L_f+1}{2L+1}\frac{\mathcal{N}_l^2 E_0^{-1}}{\pi^24^{2\Delta}\Gamma(2\Delta)(\Gamma(l+3/2))^2}4^{l+1}\pi^2 E_0^
    {-1}(l!)^2 (2E_0^{-1}E)^l \nonumber\\
    &\int_0^\infty d\omega \frac{\omega^{2l+2}\prod_{j=1}^l \left(1+\frac{\beta_E^2\omega^2}{4\pi^2j^2}\right)}{e^{\beta_E\omega}-1}.
\end{align}

Here $\beta_E$ is the inverse temperature associated with the state $\ket{E}$ in the semiclassical limit. This can be obtained using the effective form,
\begin{align}\label{eff-temp-cl}
    \beta_E = \frac{\partial S(E)}{\partial E} = \frac{\rho_S'(E)}{\rho_S(E)} \simeq \sqrt{\frac{2\pi^2}{EE_0}}.
\end{align}
Here, $S(E) = \ln\rho_S(E)$ is the associated entropy of the state $\ket{E}$, up to additive constants. The occurrence of the thermal distribution factor is not a coincidence, rather indicative of the fact that in semiclassical gravity, black holes lose their energy through thermal radiation \cite{Brown:2024ajk}. 

We can perform the $\omega$ integral by redefining $\omega = \beta_E^{-1}x$ and using \eqref{zeta}, which gives us,
\begin{align}
    \gamma_{\rm sc}(E,L) = \frac{\pi}{2}\sum_{L_f,l} &\frac{2L_f+1}{2L+1}\frac{\Gamma(2l+3)\zeta(2l+3)(l!)^2 \pi^{2l}}{\Gamma(2l+2)(\Gamma(l+3/2))^2}\left(1 + a_l\right)  \beta_E^{-1}\left(r_0\beta_E^{-1}\right)^{4l+2}.
\end{align}

We have substituted, $\mathcal{N}_l^2 E_0^{-1} = (r_0^2 E_0)^{2l+1}$. The constants $a_l$ are black hole parameter independent are coming from the polynomials of $\omega$ involving zeta functions, they vanish for $l=0$. They do not change the overall scaling of the parameters. Our semiclassical black hole has large energy compared to the coupling $E_0$ but it is still in the near-extremal regime i.e. it has a large inverse temperature compared to the horizon size. This condition can be formally written in both canonical and microcanonical ensemble as,
\begin{align}\label{semicl-near-ext-regime}
    &\beta_E\gg r_0 \nonumber\\
    & E \gg E_0, \quad \sqrt{\frac{E}{E_0}} \ll (r_0E_0)^{-1} = \frac{r_0^2}{G_N}.
\end{align}
This semiclassical but near-extremal regime ensures that the leading order contribution to the decay rate comes from the $l=0$ channel of the sum, while the $l>0$ channels are suppressed. From the selection rule, we get $L_f = L$. Hence, the leading contribution to semiclassical decay rate is given as,
\begin{align}\label{decay-semicl}
    \gamma_{\rm sc}(E) \simeq 4\zeta(3)\beta_E^{-3}r_0^2 = 4\zeta(3)r_0^2\left(\frac{E E_0}{2\pi^2}\right)^{3/2}.
\end{align}
We have dropped the $L$-label in $\gamma_{\rm sc}$ since the leading semiclassical decay rate is independent of the angular momentum of the black hole.

\subsection{Quantum decay rate}

In the quantum regime, Schwarzian states have available energies comparable to the scale $E_0$. First, we will compute the leading decay rate coming from the $l=0$ channel, which enforces $L = L_f$. Later, we will show how $l>0$ contributions are suppressed. For $l=0$, we use the asymptotic limit $E_f,E,\omega \ll E_0$. In this case, using \eqref{Sch-vertex-simple} the two-point function approximates to,
\begin{align}
    \Gamma^\Delta_{E_fE} \simeq 1.
\end{align}  
The Schwarzian density of states is given by,
\begin{align}\label{sch-dos-small-energy}
    \rho_S (E-\omega)\simeq 2\pi  \sqrt{2E_0^{-1}(E-\omega)}\Theta(E-\omega).
\end{align}
The $\Theta$-function ensures that the $\omega$ integral in the rate has a support in $(0,E)$. Using these, the leading quantum decay rate becomes,
\begin{align}\label{decay-quant}
    \gamma_{\rm qu}(E) \simeq \frac{4\sqrt{2} E_0}{15\pi}(r_0 E_0)^2 \left(\frac{E}{E_0}\right)^{5/2}.
\end{align}
This rate is also independent of the angular momentum of the state. This has only support when $E>0$.In the quantum limit, the effective temperature can be again defined from the density of states, which is given as,
\begin{align}\label{eff-temp-qu}
    \beta_E =  \frac{\rho_S'(E)}{\rho_S(E)} \simeq \frac{1}{2E}.
\end{align}

Now we will argue with the scaling of parameters that the $l>0$ contributions are suppressed. For this, let us look at the case $L = L_f$ with $1\leq l\leq 2L$. In this case, we again use $E_f, E,\omega\ll E_0$ asymptotics. Under this limit, the two-point function is again a constant $\Gamma^{\Delta}_{E_fE}\simeq (l!)^4$ and the density of states remains a polynomial \eqref{sch-dos-small-energy}. Now, defining $\omega = E y$, we get the scale dependence to be,
\begin{align*}
    \gamma_{\rm qu}^l\sim E_0(r_0E_0)^{4l+2} \left(\frac{E}{E_0}\right)^{2l+5/2}.
\end{align*}
This is parametrically small since $r_0E_0 = G_N/r_0^2 \ll 1$ and $E\ll E_0$. For $L_f > L$, the density of states vanishes. For $L_f < L$, we cannot impose $E_f\ll 1$ but the two-point function still remains $O(1)$ parameter-independent. We can again redefine $\omega$ in a similar way to get the above scale dependence of decay rate, only in this case the $y$-integral contains the $\sinh$ factor and becomes a constant, independent of $E$ and $\omega$. In this case, the scale dependence takes the form,
\begin{align*}
    \gamma_{\rm qu}^l\sim E_0(r_0E_0)^{4l+2}.
\end{align*}
For $L_f<L$, the channels require $l\geq 1$ and correspond to the release of the rotational energy $E^0_L-E^0_{L_f}$. Their rates are suppressed by at least $(r_0E_0)^6=(G_N/r_0^2)^6$ and are negligible compared to the $l=0$ channel for $E\gg E_0\,(G_N/r_0^2)^{8/5}$, which covers essentially the entire quantum regime. Hence the decay is dominated by the $l=0$ channel. Now, we are equipped to analyse a $2\to 2$ scattering process on a charged near-extremal background. This scattering event will be assumed to take place within a time scale much smaller than the decay time scale of initial and final black hole states. The above calculation shows that the decay rates of near-extremal black holes are suppressed in both the semiclassical and quantum regimes by factors of $(r_0 E_0)^2 = (G_N/r_0^2)^2$. The smallness of the decay rate justifies using the quasi-static assumption in the scattering process.

\section{$2\to 2$ scattering} \label{sec: scattering}

We are interested in a $2\to 2$ scattering process where an external particle propagates in from past infinity, interacts with the black hole, and is subsequently detected at future infinity. For this we consider an initial state of the combined system $\ket{i}\equiv\ket{E_1,L_1,M_1; \omega_1, l_1, m_1}$ at $t=-\infty$ and a final state $\ket{f}\equiv\ket{E_2,L_2,M_2; \omega_2, l_2, m_2}$ at $t=+\infty$. Here, the initial and final times are much larger compared to the interaction time, but smaller compared to the decay time scale of the external states as a result of the slow decay rates. The initial state corresponds to a Schwarzian state $\ket{\ub{E}_1}$ with a one-particle state $\ket{\ub{\omega}_1}$ of the QFT, whereas the final state corresponds to a different Schwarzian state $\ket{\ub{E}_2}$ with a one-particle state $\ket{\ub{\omega}_2}$ in the QFT. We can easiy check from the form of the interaction \eqref{int-hamil} that this process at the leading order gets contribution from second order perturbation theory. The evolution upto second order in perturbation theory is given by the operator,
\begin{align}
    U_I
&=
1
-i \int_{-\infty}^{\infty} dt\, H_I(t)
+(-i)^2
\int_{-\infty}^{\infty} dt_1
\int_{-\infty}^{t_1} dt_2\,
H_I(t_1)H_I(t_2).
\end{align}
The identity term is ignored since it is the trivial part of the amplitude, while the first-order term vanishes due to the vanishing flat space part of the amplitude owing to the fact that the interaction part is linear in creation/annihilation operators. Thus the true interaction comes from the second order piece. We define the flat space contribution as,
\begin{align}
    \mathcal{A}_F &\equiv \int_0^\infty d\omega \int_0^\infty d\omega' \bra{\ub{\omega}_2}\phi_{\ub{\omega}}(t_1)\phi_{\ub{\omega}'}(t_2) \ket{\ub{\omega}_1} \nonumber \\
    &=
u_{\omega_1}\,u_{\omega_2}^{*}
\Bigl(
e^{-i\omega_1 t_1+i\omega_2 t_2}
\,\delta_{(l m),(l_1 m_1)}
\,\delta_{(l' m'),(l_2 m_2)}
\nonumber\\ 
&\qquad\qquad
+
e^{i\omega_2 t_1-i\omega_1 t_2}
\,\delta_{(l m),(l_2 m_2)}
\,\delta_{(l' m'),(l_1 m_1)}
\Bigr).
\end{align}
Here, the two factors of the interaction Hamiltonian results in two sets of dummy variables $\ub{\omega}\equiv(\omega lm), \ub{\omega}'\equiv(\omega' l'm')$. But they get projected to particular values depending on the external particle insertions. Now we include the Schwarzian part and appropriately project the angular momenta sum,
\begin{align}
    &\mathcal{A}_{fi} = \bra{f}U_I\ket{i} \nonumber \\
&=
(-i)^2
\int_{-\infty}^{\infty} dt_1
\int_{-\infty}^{t_1} dt_2
\sum_{\substack{(l m),\\(l' m')}} \mathcal{N}_l\mathcal{N}_{l'}
\mathcal{A}_F\, 
\bra{\ub{E}_2}
e^{iH_0 t_1}
\mathcal{O}_{\ub{\Delta}}
e^{-iH_0 t_1+iH_0 t_2}
\mathcal{O}_{\ub{\Delta}'}
e^{-iH_0 t_2}
\ket{\ub{E}_1} \nonumber \\
&=
(-i)^2\mathcal{N}_{l_1}\mathcal{N}_{l_2}
u_{\ub{\omega}_1}\,u_{\ub{\omega}_2}^{*}
\int d\ub{E}\,\rho(\ub{E})
\int_{-\infty}^{\infty} dt_1
\int_{-\infty}^{t_1} dt_2
\nonumber\\
&\hspace{4em}\Bigg[
e^{-i\omega_1 t_1+i\omega_2 t_2
+i\left(E_2 + E^0_{L_2}\right) t_1-i(E + E^0_L)(t_1 -t_2)-i\left(E_1 + E^0_{L_1}\right)t_2} \mathcal{O}_{\ub{\Delta}_1}(\ub{E}_2,\ub{E})\mathcal{O}_{\ub{\Delta}_2}(\ub{E},\ub{E}_1) + \nonumber \\
&\hspace{4em}e^{i\omega_2 t_1-i\omega_1 t_2
+i\left(E_2 + E^0_{L_2}\right) t_1-i(E + E^0_L)(t_1 -t_2)-i\left(E_1 + E^0_{L_1}\right) t_2} \mathcal{O}_{\ub{\Delta}_2}(\ub{E}_2,\ub{E})\mathcal{O}_{\ub{\Delta}_1}(\ub{E},\ub{E}_1)
\Bigg].
\end{align}
Here, we have included a sum over complete set of states between the two operator insertions with labels $\ub{E}\equiv (ELM)$ and denoted $d\ub{E}\equiv \int dE \sum_{LM}$. This ensures that the free time evolution operators just become multiplicative and all the time dependencies appear in exponential form. The $t_2$-integral can be performed by introducing a small parameter $\epsilon \to 0^{+}$ that dampens the integrand at past infinity\footnote{In a short while, we will see that this $\epsilon$ parameter will be replaced by the decay rate of the internal state.}. The remaining $t_1$ integral simply gives a delta function constraint. Thus, the amplitude takes the form,
\begin{align}\label{gen-amp}
    \mathcal{A}_{fi} = &2\pi i u_{\ub{\omega}_1}\,u_{\ub{\omega}_2}^{*} \mathcal{N}_{l_1}\mathcal{N}_{l_2}
\int d\ub{E}\,\rho(\ub{E}) \delta\left(E_2 + E^0_{L_2}  +\omega_2-E_1 - E^0_{L_1}-\omega_1\right) \nonumber\\
&\Bigg[ \frac{\mathcal{O}_{\ub{\Delta}_1}(\ub{E}_2,\ub{E})\mathcal{O}_{\ub{\Delta}_2}(\ub{E},\ub{E}_1)}{E + E^0_L-E_1-E^0_{L_1}+\omega_2-i\epsilon} + \frac{\mathcal{O}_{\ub{\Delta}_2}(\ub{E}_2,\ub{E})\mathcal{O}_{\ub{\Delta}_1}(\ub{E},\ub{E}_1)}{E + E^0_L - E_1 - E^0_{L_1} - \omega_1-i\epsilon}\Bigg].
\end{align}
Here, the internal Schwarzian state remains unconstrained, while the delta function imposes an overall energy conservation, which is expected.

\subsection{Individual energy conservation and RN case}

So far, we have obtained the general form of the amplitude \eqref{gen-amp} for given initial and final states. The amplitude has poles in the upper half plane of the complex energy variable $E$. The location of the pole is slightly shifted from the resonant energy $E_* >0$ along the imaginary axis\footnote{$E_* = E_1 + E^0_{L_1} - \omega_2$ or $E_* = E_1 + E^0_{L_1} + \omega_1$ in the first and second terms respectively. The statements made in the main text are for arbitrary positive values of $E_*$.}. Since the coupling between the Schwarzian+$SU(2)$ and the QFT induces a slow decay of the energy states, we expect a pole at the location $E = E_*^R + i\gamma /2$, where $\gamma$ denotes the decay rate of the internal state and $E_*^R$ denotes the renormalised resonant energy. This ensures that the probability of the system to stay at the state $\ket{\ub{E}}$ behaves as $e^{-\gamma t}$ \cite{Weinberg1995, Schwartz2014} (See \cite{Anastopoulos:2018rms} for a review). The renormalised resonant energy can be obtained from the decay rate using the Kramers-Kronig formula,
\begin{align}\label{real-shift-resonance}
    E_*^R = E_* -\delta_\Lambda (k,L), \quad \delta_\Lambda(k,L) = \lim_{\Lambda\to\infty}\frac{1}{\pi} \mathcal{P} \int_{-\Lambda}^\Lambda dk' \frac{\gamma(k',L)}{k'-k}.
\end{align}
Here $\mathcal{P}$ denotes the principal value of the integral and $\Lambda$ denotes the UV cutoff of the theory below which our Schwarzian effective description holds. $k \equiv \sqrt{2E_0^{-1}E}$ is the momentum associated with the state. Unlike the computation of decay rate, the momentum integral is UV divergent and the result is valid only below the UV cutoff scale. This scale can be obtained by noting that the effective description and the decoupling limit of the near-extremal geometry necessarily breaks down at scales $\beta \sim r_0$. In this limit, the states are semiclassical and we can use \eqref{eff-temp-cl} to estimate the cutoff in momentum as given by $k_{\rm max} = \Lambda \sim (r_0E_0)^{-1} \gg 1$. The integral in \eqref{real-shift-resonance} can be performed by separating the quantum and semiclassical regimes where $\abs{k}\lesssim 1$ and $\abs{k} \gg 1$. The integral over the finite domain is much suppressed since $\gamma \sim \mathcal{O}(r_0 E_0)^2$ and the major contribution comes from high momentum region. In the large $\Lambda$ limit, we find two divergent pieces scaling as $\Lambda^3$ and $ k^2\Lambda$. The $k$-independent divergent term can be absorbed as a shift to the ground state energy. Since $\gamma$ itself is of order $(r_0 E_0)^2$ and $\Lambda \sim (r_0E_0)^{-1}$, the $k^2\Lambda$ piece is of order $(r_0 E_0) \sim G_N/r_0^2$. Thus we find that the real shift is much smaller compared to the bare resonant energy and can be safely ignored in estimating leading order behaviors of the quasinormal modes and Love numbers. Moreover, the real part of the shift always has an $\mathcal{O}(1)$ term while the imaginary piece starts at $\mathcal{O}(\gamma)$. With these inputs in mind, we will shift the poles in the amplitude by $-i\gamma/2$ at a given internal energy. 

Furthermore, in this work, we are interested in the background to be a near-extremal Reissner-Nordstr\"{o}m solution, and we want to extract scalar quasinormal modes and Love numbers on this background. Thus, we will compute the amplitude in a limit where the energy of the black hole and the energy of the external particle remain conserved individually. This process is thus proportional to the overlap between the initial and final states which in our convention is denoted by using \eqref{gen-amp} as,
\begin{align}\label{amp}
    \mathcal{A} = i \frac{\delta(\omega_1 - \omega_2)\delta(E_1-E_2)}{\rho(E_1)}\delta_{l_1l_2}\delta_{m_1m_2} \mathcal{M}.
\end{align}
This is true because the intermediate state in the two-point function is being traced over, multiplied with another function in the amplitude. This feature is seen in the four-point diagrams of \cite{Brown:2024ajk}. Our goal is to find the nontrivial part $\mathcal{M}$, which gets a contribution only when the energies of the initial and final states are equal i.e. $\omega_1 = \omega_2$ and $(l_1 m_1) = (l_2 m_2)$. Under these conditions, we find,
\begin{align}
    \mathcal{M} = &2\pi \abs{u_{\ub{\omega}_1}}^2\mathcal{N}_{l_1}^2\int d\ub{E} \rho(\ub{E}) \abs{\mathcal{O}_{\ub{\Delta}_1}(\ub{E}_1,\ub{E})}^2 \nonumber \\
    &\Bigg[ \frac{1}{E + E^0_L -E_1+\omega_1-i\gamma/2} + \frac{1}{E + E^0_L-E_1 - \omega_1-i\gamma/2}\Bigg].
\end{align}
Using the 3j-symbols, this simplifies to,
\begin{align}\label{amp}
    \mathcal{M} =& \frac{2\pi \abs{u_{\ub{\omega}_1}}^2\mathcal{N}_{l_1}^2E_0^{-1}}{\pi^2 2^{2\Delta_1}\Gamma(2\Delta_1)} \int_0^\infty dE\rho_S(E)\Gamma^{\Delta_1}_{EE_1} \nonumber \\
    &\Bigg[ \frac{1}{E + E^0_{l_1} -E_1+\omega_1-i\gamma/2} + \frac{1}{E + E^0_{l_1}-E_1 - \omega_1-i\gamma/2}\Bigg].
\end{align}

\section{Quasinormal modes} \label{sec: qnm}
Now, our goal is to extract the poles of the amplitude \eqref{amp-tilde}, which give us the locations of the quasinormal modes. For this, we first analyse the poles of the integrand. We will follow the redefinitions,
\begin{align}\label{mom-redef}
    &k^2 = 2E_0^{-1}E, \quad k_1^2 = 2E_0^{-1}E_1, \quad \nu^2 = l_1(l_1+1).
\end{align}
There are two types of poles of the integrand: $\omega_1$-independent or \textit{fixed poles} coming from the Schwarzian vertex in the numerator $\Gamma^{\Delta_1}_{EE_1}$ and $\omega_1$-dependent poles coming from the denominator. Our goal is to find the \textit{pole-pinching conditions} under which, these two sets of poles approach each other from two opposite sides of the contour of integration. When written in terms of the $k$ variable, our contour is the whole real axis. First, we write down the frequency-dependent poles from the denominator,
\begin{align}\label{freq-poles}
    k^2 = k_1^2 - \nu^2 \pm 2 E_0^{-1}\omega + iE_0^{-1}\gamma (k,l_1).
\end{align}
Here, the $\pm$ in front of the $\omega_1$ term corresponds to poles of the 1st and 2nd terms of \eqref{amp-tilde}. From the analysis of section \ref{sec: decay}, we see that the imaginary part $E_0^{-1}\gamma(k,l_1)$ is at least suppressed by $(r_0E_0)^2$. Thus Im$(k)$ has the same sign as Im$(k^2)$, which implies that these frequency-dependent poles are located in the upper half of the complex-$k$ plane.  Note that, if we include the real shift of the resonant pole, the value of $\omega_1$ from relation \eqref{freq-poles} will receive a small correction of $\mathcal{O}(\delta_{\Lambda})$. Since our goal is to extract only the leading order behavior of the quasinormal modes, we can safely neglect these corrections. The fixed poles of the integrand are given by the conditions,
\begin{align}\label{fixed-poles}
    k = k_{(n)} \equiv k_1 + i\lambda n, \quad \lambda = \pm 1, \, n\in \mathbb{Z} \geq l_1+1.
\end{align}
$\lambda = +1$ corresponds to poles in the upper half plane, while $\lambda = -1$ gives poles in the lower half plane. These poles come from the zeros of $\cosh(2\pi\tilde{k})-\cosh(2\pi k_1) = 0$ of \eqref{Sch-vertex-simple}. The poles for the values $n = 0,1 \cdots l$ cancel with zeros of the polynomials in the numerator.

We will analyse the pole pinching conditions i.e. when \eqref{freq-poles} and \eqref{fixed-poles} coincide, under semiclassical and quantum limits.
From the discussion above it is clear that pinching occurs only when $\lambda = -1$ as in the other case, the poles are on the same half of the complex $k$ plane and cannot pinch on any deformed contour. The pole pinching condition is given as,
\begin{align}
    \pm 2E_0^{-1}\omega_1 =  k_{(n)}^2 - k_1^2 + \nu^2 - i E_0^{-1}\gamma( k_{(n)},l_1).
\end{align}

Now we substitute $k_{(n)}$ from \eqref{fixed-poles} with $\lambda = -1$ and extract the leading behaviour of the low-lying QNMs for $n,l_1 \mathcal{O}(1)$. In this limit, the decay factor and in turn the real shift remain suppressed in both semiclassical and quantum regimes. The damped modes come from the emission channel and they are given by,
\begin{align}\label{QNM}
    \boxed{\omega =\left( l(l+1) - n^2\right)\frac{E_0}{2} - i\sqrt{2E_0 E_1} n, \quad n\geq l+1} .
\end{align}
We have dropped the \lq 1' subscripts. Thus we find that the leading behaviour of the low-lying quasinormal modes, expressed in terms of available energies of the near-extremal black holes, have the same structures in the semiclassical and quantum regimes. Since $E_0 = G_N/r_0^3$, we find that the imaginary part dominates over the real part in the semiclassical regime $E_1 \gg E_0$. The semiclassical result can also be written in the familiar form of zero-damped modes in the literature \cite{Hod:2008se, Zimmerman:2015trm},
\begin{align}\label{QNM-cl-lead}
    \boxed{\omega_{\rm sc} =  \left(l(l+1) - n^2\right)\frac{G_N}{2r_0^3} - 2\pi i T n, \quad n\geq l+1}. 
\end{align}
Here, we have used the semiclassical average temperature from \eqref{eff-temp-cl}. In contrast, in the quantum limit, the imaginary part can be suppressed compared to the real part. In terms of the quantum average temperature, the QNMs take the form,
\begin{align}\label{QNM-qu-lead}
    \boxed{\omega_{\rm qu} =  \left(l(l+1) - n^2\right)\frac{G_N}{2r_0^3} - i\sqrt{\frac{G_N T}{r_0^3}}n}.
\end{align}

\section{Static Love numbers}\label{sec: Love}
A key discriminant of compact objects in binary coalescence is the static Love number~\cite{10.1098/rspa.1909.0008}. The static Love number characterizes the tidal deformability under a slowly varying external tidal field. It is known to vanish for classical black holes and may receive nonvanishing quantum corrections. The approach presented in this paper provides an excellent opportunity to constrain the Love numbers in the quantum and semiclassical regimes. First, we give a lightning review of Love numbers.

\subsection{A brief review}
The Love number is ideally regarded as a property of the EFT, or as a Wilson coefficient in an EFT constructed to capture the dynamics of a compact object in the long-wavelength regime. This regime is naturally realized during the binary inspiral. Generally, we replace the compact object with an effective worldline description. For a compact object that interacts gravitationally with a scalar field, a suitable worldline action may be written as~\cite{Elkhidir:2026lgp}
\begin{equation}
\label{eq:FullAction}
\begin{split}
    S &= -M\int d\tau
    + \int d^4x\,\sqrt{-g}
    \left[
    \frac{1}{2}\,g^{\mu\nu}
    \partial_\mu\phi\,\partial_\nu\phi
    - \frac{2}{\kappa^2}R
    \right]
    \\
    &\quad
    + \frac{1}{2}
    \sum_{l=0}^{\infty}\sum_{n=0}^{\infty}
    \left(2\pi\Lambda_{l}^{(n)}\right)
    \kappa^2\frac{M}{2}\int d\tau\,
    \big[
    \partial_{\parallel}^n
    \partial_{\langle\nu\rangle_{l}\perp}
    \phi(z(\tau))
    \big]
    \big[
    \partial_{\parallel}^n
    \partial_{\perp}^{\langle\nu\rangle_{l}}
    \phi(z(\tau))
    \big] \, .
\end{split}
\end{equation}
where $\kappa$ is the graviton coupling defined as
\begin{equation}
    \kappa^2 \equiv 32\pi G_N
    = \frac{32\pi}{M_{\rm pl}^2}
    = 32\pi l_{\rm pl}^2 \, .
\end{equation}

The first line of Eq.~\eqref{eq:FullAction} is the sum of the point-particle action, the bulk scalar kinetic term, and the Einstein-Hilbert action. In the point-particle action, the proper time is defined by
\begin{equation}
    d\tau = \sqrt{g_{\mu\nu}dz^\mu dz^\nu} \, .
\end{equation}
The second line contains the operators describing the body's scalar tidal response. The symbol $\partial_{\parallel}$ denotes the derivative parallel to the four velocity $v^\mu=dz^\mu/d\tau$ of the worldline, while $\partial_{\langle\nu\rangle_{l}\perp}$ denotes the symmetric trace-free product of $l$ derivatives orthogonal to the four velocity. More precisely,
\begin{equation}
\label{eq:STFnotationv}
\begin{split}
    \partial_{\parallel}
    &\equiv v^\mu\partial_\mu \, ,
    \\
    \partial_{\langle\mu\rangle_{l}\perp}
    &\equiv
    \text{STF}
    \left[
    \prod_{i=1}^{l}
    \mathbb{P}_{\mu_i}^{\ \nu_i}
    \partial_{\nu_i}
    \right]
    =
    \text{STF}
    \left[
    \prod_{i=1}^{l}
    \left(
    \delta_{\mu_i}^{\nu_i}
    -v_{\mu_i}v^{\nu_i}
    \right)
    \partial_{\nu_i}
    \right] \, .
\end{split}
\end{equation}
The projector $\mathbb{P}_{\mu}^{\ \nu}$ projects orthogonally to the worldline, onto the spatial slice in the instantaneous rest frame of the body. STF denotes the symmetric trace-free projection. We can now restrict to static Love numbers by considering only terms with $n=0$, yielding
\begin{equation}
\label{eq:StaticAction}
\begin{split}
    S &= -M\int d\tau
    + \int d^4x\,\sqrt{-g}
    \left[
    \frac{1}{2}\,g^{\mu\nu}
    \partial_\mu\phi\,\partial_\nu\phi
    - \frac{2}{\kappa^2}R
    \right]
    \\
    &\quad
    + \frac{1}{2}
    \sum_{l=0}^{\infty}
    \left(2\pi\Lambda_{l}^{(0)}\right)
    \kappa^2\frac{M}{2}\int d\tau\,
    \big[
    \partial_{\langle\nu\rangle_{l}\perp}
    \phi(z(\tau))
    \big]
    \big[
    \partial_{\perp}^{\langle\nu\rangle_{l}}
    \phi(z(\tau))
    \big] \, .
\end{split}
\end{equation}

The coefficients $\Lambda_l^{(0)}$ are the dimensionful Wilson coefficients describing the static tidal response. The factor of $2\pi$ in our convention is included to account for the normalization of the spherical-wave states used below. We will introduce the corresponding dimensionless Love numbers after performing the spherical-wave decomposition.

We can study the contribution of these operators to the scattering of a scalar wave off the compact object. At leading order in the tidal response, the plane-wave amplitude is
\begin{equation}
    i\mathcal{A}^{\rm Love}_{\mathbf{k}\rightarrow\mathbf{l}}
    =
    i\sum_{L=0}^{\infty}
    \left(2\pi\Lambda_L^{(0)}\right)
    \kappa^2\frac{M}{2}\,
    k_{\langle\nu_1\cdots\nu_L\rangle\perp}
    l^{\langle\nu_1\cdots\nu_L\rangle}_{\perp} \, .
\end{equation}
The factor of $1/2$ in the quadratic worldline operator is cancelled by the two contractions of the identical scalar fields.

In the rest frame of the compact object,
\begin{equation}
    k_{\langle\nu_1\cdots\nu_L\rangle\perp}
    l^{\langle\nu_1\cdots\nu_L\rangle}_{\perp}
    =
    |\mathbf{k}|^L|\mathbf{l}|^L
    \hat{k}_{\langle i_1\cdots i_L\rangle}
    \hat{l}^{\langle i_1\cdots i_L\rangle} \, .
\end{equation}
Using the STF identity
\begin{equation}
    \hat{k}_{\langle i_1\cdots i_L\rangle}
    \hat{l}^{\langle i_1\cdots i_L\rangle}
    =
    \frac{L!}{(2L-1)!!}
    P_L(\hat{\mathbf{k}}\cdot\hat{\mathbf{l}})
    =
    \frac{4\pi L!}{(2L+1)!!}
    \sum_{q=-L}^{L}
    Y_{Lq}(\hat{\mathbf{k}})
    Y^*_{Lq}(\hat{\mathbf{l}}) \, ,
\end{equation}
we can convert the amplitude to the spherical-wave basis.

We follow the construction of Appendix B of Ref.~\cite{Saketh:2023bul}, with a slightly different normalization of the spherical-wave states. In Ref.~\cite{Saketh:2023bul}, the states obtained by setting $h=0$ satisfy
\begin{equation}
    {}_{\rm Ref}\langle\omega,l,m
    |\omega',l',m'\rangle_{\rm Ref}
    =
    2\pi\,
    \delta(\omega-\omega')
    \delta_{ll'}
    \delta_{mm'} \, .
\end{equation}
Here we instead use the normalization
\begin{equation}
    \langle\omega,l,m
    |\omega',l',m'\rangle
    =
    \delta(\omega-\omega')
    \delta_{ll'}
    \delta_{mm'} \, ,
\end{equation}
which agrees with the normalization of the free QFT modes used in the Schwarzian theory. Thus,
\begin{equation}
    |\omega,l,m\rangle
    =
    \frac{1}{\sqrt{2\pi}}\,
    |\omega,l,m\rangle_{\rm Ref} \, .
\end{equation}
The scalar plane-wave to spherical-wave overlap in our convention is consequently
\begin{equation}
    \langle\omega,l,m|\mathbf{k}\rangle
    =
    (2\pi)^{3/2}
    \sqrt{\frac{2}{\omega}}\,
    (-1)^m
    \delta(\omega-|\mathbf{k}|)
    Y_{l,-m}(\hat{\mathbf{k}}) \, .
\end{equation}

The spherical-basis amplitude is therefore
\begin{align}
    i\mathcal{M}^{\rm Love}_{\omega l  m}
    &=
    \int
    \frac{d^3\mathbf{k}\,d^3\mathbf{l}}
    {(2\pi)^6\,4|\mathbf{k}||\mathbf{l}|}
    \langle\omega,l,m|\mathbf{l}\rangle\,
    i\mathcal{A}^{\rm Love}_{\mathbf{k}\rightarrow\mathbf{l}}\,
    \langle\mathbf{k}|\omega,l,m\rangle
    \nonumber\\
    &=
    \frac{1}{16\pi^3\omega}
    \int
    d^3\mathbf{k}\,d^3\mathbf{l}\,
    \frac{
    \delta(\omega-|\mathbf{k}|)
    \delta(\omega-|\mathbf{l}|)
    }{
    |\mathbf{k}||\mathbf{l}|
    }\,
    Y^*_{l,-m}(\hat{\mathbf{k}})
    Y_{l,-m}(\hat{\mathbf{l}})
    \,
    i\mathcal{A}^{\rm Love}_{\mathbf{k}\rightarrow\mathbf{l}} \, .
\end{align}
Relative to the normalization of Ref.~\cite{Saketh:2023bul}, the two external spherical-wave states therefore contribute an overall factor of $1/(2\pi)$. This factor precisely cancels the factor of $2\pi$ included in the worldline vertex.

Substituting the plane-wave amplitude, the angular integral becomes
\begin{align}
    &\int d\Omega_{\mathbf{k}}\,d\Omega_{\mathbf{l}}\,
    Y^*_{l,-m}(\hat{\mathbf{k}})
    Y_{l,-m}(\hat{\mathbf{l}})
    \hat{k}_{\langle i_1\cdots i_L\rangle}
    \hat{l}^{\langle i_1\cdots i_L\rangle}
    \nonumber\\
    &\qquad
    =
    \frac{4\pi L!}{(2L+1)!!}\,
    \delta_{Ll} \, .
\end{align}
Thus only the $L=l$ Love coefficient contributes.

The radial integrals are
\begin{equation}
    \int_0^\infty dk\,
    k^{L+1}\delta(\omega-k)
    =
    \omega^{L+1} \, .
\end{equation}
Together with the overall factor of $1/\omega$, this gives
\begin{align}
    i\mathcal{M}^{\rm Love}_{\omega l  m}
    &=
    i\frac{1}{16\pi^3\omega}
    \left(
    2\pi\Lambda_l^{(0)}
    \kappa^2\frac{M}{2}
    \right)
    \frac{4\pi l!}{(2l+1)!!}
    \omega^{2l+2}
    \nonumber\\
    &=
    i\,\Lambda_l^{(0)}
    \frac{\kappa^2 M}{4\pi}
    \frac{l!}{(2l+1)!!}
    \omega^{2l+1} \, .
\end{align}
The result is independent of $m$, as expected from spherical symmetry.

Therefore,
\begin{equation}
    i\mathcal{M}^{\rm Love}_{\omega l m}
    =
    i\,\Lambda_l^{(0)}
    \frac{\kappa^2 M}{4\pi}
    \frac{l!}{(2l+1)!!}
    \omega^{2l+1} \, ,
\end{equation}

For the near-extremal black holes considered here, the relevant mass scale is provided by the extremal radius
\begin{equation}
    r_0 = G_N M \, .
\end{equation}
We define the dimensionless static Love number as
\begin{equation}
    \lambda_l^{(0)}
    \equiv
    \Lambda_l^{(0)}r_0^{-2l} \, .
\end{equation}
Using
\begin{equation}
    \kappa^2 M = 32\pi r_0 \, ,
\end{equation}
the spherical-basis amplitude becomes
\begin{equation}\label{eq:tomatch}
    i\mathcal{M}^{\rm Love}_{\omega l m}
    =
    i\,8\,\lambda_l^{(0)}
    \frac{l!}{(2l+1)!!}
    (r_0\omega)^{2l+1} \, ,
\end{equation}

Below, we will match this with the corresponding scalar-scattering amplitude in the Schwarzian theory to fix the static Love numbers. However, it is important to mention the assumptions involved in this analysis.

The Schwarzian theory is expected to describe well the near-horizon region of near-extremal black holes. However, the full problem of scalar scattering off a compact object also involves diagrams that do not probe the near-horizon physics, such as scattering off the background curvature. The total scattering quantity thus emerges from the combination of the two, and it is this quantity that must be matched with its counterpart in an appropriately quantum-corrected worldline EFT to obtain Wilson coefficients such as the static Love number. This is known to be a very challenging problem, although important progress has been made in the last couple of years for Schwarzschild black holes~\cite{Caron-Huot:2025tlq,Correia:2026utp,Ivanov:2026icp,Ivanov:2024sds,Bautista:2026fcp,Brunello:2026rdk, Brunello:2026lzf} and spinless neutron stars~\cite{Saketh:2024juq,Saketh:2026trm}. These works do not generally include quantum corrections to the near-horizon physics.

However, it has also been observed and exploited in many previous works that the tidal response seems to arise primarily from the near-zone physics~\cite{Ivanov:2022qqt,Saketh:2024juq,Chakraborty:2026dox,Creci:2021rkz}. In fact, the scattering phase computed in black-hole perturbation theory naturally seems to permit such a separation to some degree~\cite{Ivanov:2022qqt}. This allows us to constrain some of the Wilson coefficients, such as the static Love number, without including far-zone contributions to the process. Following that logic, in this work we assume that the scattering amplitude computed in the Schwarzian theory in the near-horizon region encodes the leading static tidal deformability and may thus be directly compared with Eq.~\eqref{eq:tomatch}. 

In worldline EFT, the scattering amplitude used for fixing linear tidal response assumes that the black hole has not meaningfully changed under the process, and thus we compute correlation functions of fields in a black-hole background in some sense. Thus, the initial and final scattering states of the black hole are identical, while the wave can change momentum. Thus, in the Schwarzian description, we compute amplitudes with the same state of the black hole in initial and final states. This is justified even if the states can decay because the decay rate is parametrically suppressed as explained before.
\subsection{Conservative response}
Now, we will compute the conservative response characterised by static Love numbers. These can be done by finding the real part of the amplitude \eqref{amp}. First, we drop the \lq 1' subscripts from the external scalar mode in \eqref{amp} and make a change of variable from energy to momenta. We again use \eqref{mom-redef}, which gives,
\begin{align}\label{amp-tilde}
        &\mathcal{M}(\omega) = \frac{2\pi \abs{u_{\ub{\omega}}}^2\mathcal{N}_{l}^2E_0^{-1}}{\pi^2 2^{2\Delta}\Gamma(2\Delta)} \tilde{\mathcal{M}}(\omega) \\
        & \tilde{\mathcal{M}}(\omega) = \int_{-\infty}^{\infty} dk \, k \sinh(2\pi k) \Gamma^{\Delta}_{kk_1}  \Bigg[ \frac{1}{k^2 + \nu^2 -k_1^2+2E_0^{-1}\omega-i\gamma E_0^{-1}} + (\omega \to -\omega)\Bigg].
\end{align}
Now, we determine the prefactor in the expression of,
\begin{align}\label{prefactor}
    \frac{2\pi \abs{u_{\ub{\omega}}}^2\mathcal{N}_{l}^2E_0^{-1}}{\pi^2 2^{2\Delta}\Gamma(2\Delta)} = \frac{2}{2^{2\Delta}\pi\Gamma(2\Delta)\Gamma(l+3/2)^2} r_0^{2l+1} E_0^{2\Delta -1}(r_0\omega)^{2l+1}.
\end{align}
This already shows the relevant scaling in terms of $(r_0\omega)^{2l+1}$. Thus, we will evaluate the real part of the amplitude $\tilde{\mathcal{M}}(\omega)$ in the $\omega\to 0$ limit and it will precisely be proportional to the static Love numbers. Let us denote it by the integral,
\begin{align}\label{real-part}
    \frac{1}{2}{\rm Re}\tilde{\mathcal{M}}(0) \equiv I_\Delta, \quad I_\Delta = \mathcal{P}\int_{-\infty}^{\infty} dk \, \frac{k \sinh(2\pi k) \Gamma^{\Delta}_{kk_1}}{k^2 + \nu^2 -k_1^2}.
\end{align}
Here, we have used the identity \eqref{PV} and $\mathcal{P}$ denotes the principal value of the integral. Also, the Schwarzian vertex factor \eqref{Sch-vertex} is naturally a function of momenta and it is given as,
\begin{align}
    \Gamma^{\Delta}_{kk_1} = \prod_{\pm}\Gamma(\Delta\pm ik\pm ik_1).
\end{align}

Now, we focus on the large-$k$ behavior of the integrand using \eqref{Gamma-large-y},
\begin{align}\label{large-k}
    \frac{\abs{k} e^{2\pi\abs{k}}}{k^2} \abs{k}^{4\Delta - 2} e^{-2\pi\abs{k}} \sim \abs{k}^{4\Delta - 3}.
\end{align}
We can readily see the integrand dies off at large $\abs{k}$ for values Re$\Delta < 1/2$. The values of $\Delta$ we are interested in are $\Delta = l+1 \geq 1$, where the integral has UV divergences. One way to regulate the divergence is by introducing the UV cutoff $k_{\rm max}$.

However, for convenience, we will evaluate the integration by analytic continuation of $I_\Delta$ in the open domain $0 < {\rm Re}\Delta < 1/2$, where it perfectly converges. Let us briefly summarise the process of analytic continuation. Firstly, we treat the conformal dimension in Schwarzian theory to be lying in the open domain $0 < {\rm Re}\Delta < 1/2$ of the complex $\Delta$ plane while keeping the angular momentum $l$ of the external scalar fixed. Thus, we only complexify the conformal dimension $\Delta$, a characteristic of the effective theory, but $\nu^2 = l(l+1)$ and the external legs of the scalar are kept fixed. Next, we compute the integral in the amplitude using a semicircular contour in the complex $k$-plane, with a technique such that the contributions of the gamma function poles vanish.

For this, we add an odd function to the integrand, which has zero contribution on the real line. Thus, the integral on the real line remains unchanged. This is given by,
\begin{align}\label{I-Delta}
    &I_{\Delta} = \mathcal{P}\int_{-\infty}^{\infty} dk \, \frac{k  \Gamma^{\Delta}_{kk_1}}{k^2 + \nu^2 -k_1^2}g(k), \quad \\
    &g(k) = \sinh(2\pi k) - \frac{i}{\sin(2\pi\Delta)}\left[\cosh(2\pi k_1) - \cos(2\pi\Delta)\cosh(2\pi k)\right].
\end{align}
It is evident from the form that the additional even part of $g(k)$ makes the full integrand odd and its contribution vanishes. We will evaluate $I_\Delta$ for $0 < {\rm Re}\Delta < 1/2$ using a semicircular contour in the upper half of the complex $k$-plane. Owing to \eqref{large-k}, at large $\abs{k}$ the integrand dies off in this regime and this behavior remains unchanged even in presence of the even part of $g(k)$. We find that the poles of $\Gamma^\Delta_{kk_1}$ in the UHP are located at,
\begin{align}
    k = \pm k_1 + i(\Delta + n), \quad n\in \mathbb{Z} \geq 0.
\end{align}
These are simple poles. Now, we can easily check that, the function $g(k)$ vanishes at the location of these poles,
\begin{align}
    &g(\pm k_1 + i(\Delta + n)) \nonumber \\
    =&\,  i\sin(2\pi (\Delta\pm ik_1)) - \frac{i}{\sin(2\pi\Delta)}\left[\cosh(2\pi k_1) - \cos(2\pi\Delta)\cos(2\pi (\Delta\pm ik_1))\right] = 0
\end{align}
Thus, in the contour integral, these poles will give vanishing contributions. This is the most important trick in evaluating the integral and justifies our choice of $g(k)$. We will compute the integral for zero and nonzero threshold energies and extract static Love numbers in the classical, semiclassical, and quantum regimes. 

\subsection{Zero threshold: Classical and Quantum $l=0$ }

We first consider the case $\nu = 0$, where the threshold energy above extremality is zero. This covers firstly the $l=0$ response in both classical and quantum regimes. Moreover, the strictly classical physics $(k_1 \gg 1)$ is recovered at $E_0 \to 0$ since in classical gravity, we do not expect a threshold in the spectrum of near-extremal black holes. Hence this case also covers all $l \neq 0$ classical response. In this case, the integral $I_\Delta$ \eqref{I-Delta} becomes,
\begin{align}
    &I_{\Delta} = \mathcal{P}\int_{-\infty}^{\infty} dk \, \frac{k  \Gamma^{\Delta}_{kk_1}}{k^2  -k_1^2}g(k).
\end{align}
The integrand has poles on the real axis at values $k = \pm k_1$. Thus, to extract the Principal value, we go around these poles along two small semicircles centered at $k = \pm k_1$. The closed contour now only contains the simple poles coming from the Gamma functions, however, since the integrand vanishes at these locations, the residues are zero. Thus, we get,
\begin{align}
    I_\Delta = - \int_{C_{-}}dk \, \frac{k  \Gamma^{\Delta}_{kk_1}}{k^2  -k_1^2}g(k)  - \int_{C_{+}}dk \, \frac{k  \Gamma^{\Delta}_{kk_1}}{k^2  -k_1^2}g(k),
\end{align}
here $C_{\pm}$ denote the small semicircular contours centered at $\pm k_1$. On these contours, we have $k = \pm k_1 + r e^{i\varphi}$ for $r\to 0$ and $\pi \leq \varphi \leq 0$. Due to the reflection symmetry, only the even part of $g(k)$ contributes to the sum. Thus, we get,
\begin{align}\label{I-Delta-real-pole}
    I_\Delta &= \frac{i\pi}{2} \Gamma^\Delta_{k_1k_1}(g(k_1)+g(-k_1)).
\end{align}
Evidently, only the even part of $g(k)$ contributes to the integral. Simplifying this, we find,
\begin{align}
    I_\Delta &=i\pi \Gamma^\Delta_{k_1k_1} \frac{(-i)}{\sin (2\pi\Delta)} \left[\cosh(2\pi k_1) - \cos(2\pi\Delta)\cosh(2\pi k_1)\right] \nonumber \\
    &= \pi \Gamma^\Delta_{k_1k_1} \cosh(2\pi k_1) \tan(\pi\Delta).
\end{align}
This result is analytic in the open domain $0 < {\rm Re}\Delta < 1/2$ and can be extended to the values of our interest $\Delta = l+1$. We find that in these values $\tan(\pi(l+1)) = 0$, thus the conservative response vanishes,
\begin{align}
    \boxed{{\rm Re}\mathcal{M}(0) \propto \tan(\pi(l+1)) = 0}.
\end{align}
Hence, the classical and quantum $l=0$ static Love numbers are identically zero,
\begin{align}\label{Love-cl}
    \boxed{\lambda^{(0)}_{l,{\rm cl}} = 0, \quad \lambda^{(0)}_{0,{\rm qu}} = 0}.
\end{align}
The vanishing of $l=0$ is a consistent with the shift symmetry of massless scalars~\cite{Parra-Martinez:2025bcu}. The vanishing of classical static Love numbers also supports the predictions of GR and together these two features serve as a very important consistency check of our computations.

\subsection{Non-zero threshold}
Now, we will evaluate the integral for the non-zero threshold energy i.e., for the fixed value $\nu = \sqrt{l(l+1)}>0$. In this case, the UHP poles of the denominator are located at the zeros of $k^2 - k_1^2 + \nu^2 = 0$,
\begin{align}
    k = \pm k_0, \quad k_0 = \sqrt{k_1^2 - \nu^2}, \quad {\rm when} \, k_1 >\nu, \\
    k = + i\mu, \quad \mu = \sqrt{\nu^2 - k_1^2}, \quad {\rm when} \, k_1 <\nu.
\end{align}
First, we consider the case $k_1 > \nu$ for which the poles are located on the real axis. In this case, we again deform the contour close to the poles around small semicircles. Again, the residue at the poles of the gamma functions are zero such that the contour integral vanishes. Only the small semicircles contribute to the integral of our interest. We find that the result in this regime can be obtained by replacing the pole location $k_1$ to $k_0$ in \eqref{I-Delta-real-pole},
\begin{align}
    I^{k_1>\nu}_\Delta &= \frac{i\pi}{2} \Gamma^\Delta_{k_0 k_1}(g(k_0)+g(-k_0)) = i\pi \Gamma^\Delta_{k_0 k_1} g_{\rm even} (k_0).
\end{align}
Here $g_{\rm even}$ denotes the even part of the function $g$. We now extend this analytic result from the domain $0 < {\rm Re}\Delta < 1/2$ and to the values of our interest $\Delta = l+1$. In this limit, we find the behavior of the functions,
\begin{align}
    &\Gamma^\Delta_{k_0 k_1} \simeq \Gamma^{l+1}_{k_0 k_1} + \left(\partial_\Delta \Gamma^\Delta_{k_0 k_1}\right)_{\Delta\to l+1} \delta + \mathcal{O}(\delta^2), \\
    & g_{\rm even} (k_0) \simeq \frac{-i}{2\pi\delta} (\cosh 2\pi k_1 - \cosh 2\pi k_0) + \mathcal{O}(\delta).
\end{align}
Here, $\delta = \Delta - l-1 \to 0$. Using \eqref{Sch-vertex-simple}, we find that,
\begin{align}
    \Gamma^{l+1}_{k_0 k_1}  = \frac{2 P(k_1^2)}{\cosh 2\pi k_1 - \cosh 2\pi k_0},
\end{align}
where $P(k_1^2)$ is a smooth function of $k_1^2$ i.e. the $\bra{E_1}H_0\ket{E_1}$ eigenvalues,
\begin{align}\label{P-func}
    P(k_1^2) = \pi^2\nu^2\prod_{j=1}^l \left[j^4 + \nu^4 + 2j^2 (2k_1^2 - \nu^2)\right].
\end{align}
We also define,
\begin{align}\label{Q-func}
    Q(k_1) = \frac{1}{2}\left(\partial_\Delta\Gamma^\Delta_{k_0 k_1}\right)_{\Delta\to l+1}(\cosh 2\pi k_1 - \cosh 2\pi k_0).
\end{align}
Now, the integral is given as,
\begin{align}\label{real-pole-int}
    I^{k_1>\nu}_{\Delta\to l+1} = \frac{P(k_1^2)}{\delta} + Q(k_1).
\end{align}

Next, we consider the opposite case $k_1 < \nu$, where there is no pole on the real axis but one pole on the imaginary line. In this case, we get,
\begin{align}
    I^{k_1<\nu}_\Delta &= \mathcal{P}\left(i\pi \Gamma^\Delta_{i\mu,k_1} g(i\mu)\right).
\end{align}
Using $\mu = -ik_0 >0$, we obtain the simple relation,
\begin{align}
    I^{k_1<\nu}_\Delta &= I^{k_1>\nu}_\Delta - \frac{2\pi P(k_1^2)\sin 2\pi \mu}{\cosh2\pi k_1 - \cos 2\pi\mu}.
\end{align}
Here, the last term has appeared from the odd part of the function $g$. Using the identity \eqref{gamma-prod-l}, we can easily show that both the cases $k_1>\nu$ and $k_1 < \nu$ can be collectively expressed in a simple form,
\begin{align}\label{contd-int}
    &I_{\Delta\to l+1} = \frac{P(k_1^2)}{\delta} + Q(k_1) - \frac{2\pi a P(k_1^2)\sin 2\pi \mu}{\cosh2\pi k_1 - \cos 2\pi\mu}.
\end{align}
Here, $a = 0$ for $k_1 > \nu$ and $a = 1$ for $k_1 <\nu$. Now, we can find the real part of the amplitude \eqref{amp-tilde} and \eqref{real-part} using \eqref{prefactor},
\begin{align}\label{bare-re-amp}
    {\rm Re} \mathcal{M}(0) = &\frac{4(M_0^{-1}\omega)^{2l+1}}{2^{4l+4}\pi\Gamma(2l+2)\Gamma(l+3/2)^2} \nonumber\\
    &\times\left(\frac{P(k_1^2)}{\delta} + 2P(k_1^2)\log \frac{E_0}{2e^{\psi(2l+2)}} + Q (k_1) - \frac{2\pi a P(k_1^2)\sin 2\pi \mu}{\cosh2\pi k_1 - \cos 2\pi\mu}\right).
\end{align}
Here, we have used $E_0 = G_N/r_0^3$ and $M_0 = G_N^{-1}r_0$ is the extremal mass. The $\log E_0$ term arises because the analytic continuation $\Delta = l + 1 + \delta$ leaves an extra factor of $E_0^{2\delta}$ in the amplitude.

\subsubsection{Counterterms}
The divergent piece of order $\delta^{-1}$ in the amplitude \eqref{bare-re-amp} can be removed by adding a local time-dependent counterterm in the Hamiltonian. Unlike the source-operator interaction, this counterterm is quadratic in the quantum field. We define this as,
\begin{align}
    H_{\rm ct} (t) = \sum_{lm} \hat\Lambda_l(H_0)\int_0^\infty d\omega \int_0^\infty d\omega' :\phi_{\omega lm}(t)\phi_{\omega' lm}(t) :
\end{align}
Here the coefficient $\hat\Lambda_l(H_0)$ is a local function of the free Hamiltonian of the Schwarzian + $SU(2)$ theory and it is designed to be diagonal in the eigenbasis of $H_0$. Since this counterterm is already second order in the far-field, it contributes to the first-order perturbation term in the Dyson series. Also, it is of the same order in $\phi$ as our amplitude \eqref{amp-tilde}. Now, we find the contribution of this term to the transition,
\begin{align}
   -i\int_{-\infty}^{\infty} dt\  \bra{f}H_{\rm ct} (t)\ket{i} = \langle f|i\rangle \sum_{lm} \hat\lambda_l(E_1) \left(\frac{\omega}{M_0}\right)^{2l+1}. 
\end{align}
Here, we have absorbed an additional factor of $(-4\pi i M_0^{2l+1})$ into the operator, and $E_1$ is the eigenvalue of $H_0$ on the initial state $\ket{E_1,0,0}$. The initial and final states are taken as the ones given in section \ref{sec: scattering}. This counterterm precisely has the required scaling of static Love numbers and by choosing its appropriate form, we can absorb the $\delta\to 0$ divergence of the conservative response \eqref{bare-re-amp}. The renormalised response is given by,
\begin{align}
    {\rm Re}\mathcal{M}_R(0) = {\rm Re}\mathcal{M}(0)  + \hat\lambda_l(E_1).
\end{align}
The required operator is given by,
\begin{align}
    & \hat\lambda_l(H_0) = -\frac{4\varepsilon^{2\delta}}{2^{4l+4}\pi\Gamma(2l+2)\Gamma(l+3/2)^2} \left(\frac{1}{\delta} P\left(\frac{2H_0}{E_0}\right) +\hat{\lambda}^{\rm ren}_l(H_0) \right).
\end{align}
Here, $P\left(\frac{2H_0}{E_0}\right)$ is the operator valued polynomial \eqref{P-func} and $\hat{\lambda}^{\rm ren}_l(H_0)$ is a scheme-dependent finite piece, which can be a local function of $H_0$. Since for $\Delta = l + 1 + \delta$ the amplitude \eqref{amp-tilde} carries an
additional factor of $E_0^{2\delta}$, the counterterm coefficient must carry a
compensating factor $\varepsilon^{2\delta}$, where $\varepsilon$ is an arbitrary
energy scale. Redefining $\varepsilon \to 2e^{\psi(2l+2)}\varepsilon$, the finite part of the renormalised amplitude then depends on
$\log(E_0/\varepsilon)$, and the dependence on $\varepsilon$ reflects the choice
of renormalisation scheme. With this choice of counterterm, the renormalised conservative response can be written as,
\begin{equation}\label{renorm-amp}
\boxed{
\begin{aligned}
{\rm Re}\,\mathcal{M}_R(0) = &\frac{4\,(M_0^{-1}\omega)^{2l+1}}{2^{4l+4}\pi\,\Gamma(2l+2)\,\Gamma(l+3/2)^2} \\
&\times \bigg( Q(k_1) + 2P(k_1^2)\log\frac{E_0}{\varepsilon} - \frac{2\pi a\, P(k_1^2)\sin 2\pi\mu}{\cosh 2\pi k_1 - \cos 2\pi\mu} + \hat{\lambda}^{\rm ren}_l(k_1^2) \bigg)
\end{aligned}
}
\end{equation}
Here, $\mu = \sqrt{\nu^2-k_1^2}$ and $a = 0/1$ for $k_1 > \nu$ or $k_1 < \nu$. Now, comparing with \eqref{eq:tomatch}, we find the quantum-corrected static Love numbers for arbitrary $k_1$ in the near-extremal regime,
\begin{align}
\label{all-Love}
\boxed{
\begin{aligned}
\lambda^{(0)}_{l,\mathrm{near-ext}}
&=
\frac{
(2l+1)!!\,\left(G_N/r_0^2\right)^{2l+1}
}{
2^{4l+5}\pi\,l!\,
\Gamma(2l+2)\,
\Gamma(l+3/2)^2
}
\\[2mm]
&\times
\left[
Q\!(k_1)
+ 2P(k_1^2)\log\frac{E_0}{\varepsilon}-\frac{
2\pi a\,P(k_1^2)\sin(2\pi\mu)
}{
\cosh(2\pi k_1)-\cos(2\pi\mu)
}
+\hat{\lambda}^{\mathrm{ren}}_l
\!\left(k_1^2\right)
\right].
\end{aligned}
}
\end{align}
This gives us a general expression for static Love number at both the semiclassical and quantum regimes. We see that the exact Love number is dependent on our renormalization scheme. But it also contains non-trivial energy dependence via $Q\left(k_1\right)$ which encodes the physical information of the tidal response. Below we discuss the semiclassical and quantum behaviour of this expression.

\subsubsection{Semiclassical vs Quantum Love}
First, we consider the semiclassical limit $k_1 \gg \nu$, where $a=0$. Under this approximation, we obtain,
\begin{align}
    Q(k_1) + 2P(k_1^2)\log\frac{E_0}{\varepsilon}
&= \nu^2 k_1^{2l}\Big[\alpha_l\log k_1+\gamma_l\Big] + \nu^4 k_1^{2l-2}\Big[\alpha'_l\log k_1+\gamma'_l\Big] \nonumber \\
&+\mathcal{O}\!\left(\nu^6 k_1^{2l-4},\,\nu^2 k_1^{2l-2}\right).
\end{align}
Now, we absorb the coefficient,$$c_l = \frac{(2l+1)!!}
{2^{4l+5}\pi\,l!\,\Gamma(2l+2)\,\Gamma(l+3/2)^2}$$
into the above constants $(c_l\alpha_l,c_l \beta_l \cdots) \equiv (\alpha_l,\beta_l \cdots)$. These are given as,
\begin{align*}
   &\alpha_l = 2\mathcal{B}_l, \quad \beta_l = \mathcal{B}_l, \quad
\gamma_l = 2\mathcal{B}_l\Big[\log\frac{2E_0}{\varepsilon} + \psi(l+1)\Big], \quad
\mathcal{B}_l \equiv c_l\, 2^{2l}\pi^2 (l!)^2, \\
&\alpha'_l = -l\,\mathcal{B}_l, \quad \beta'_l = -\frac{l}{2}\mathcal{B}_l, \quad
\gamma'_l = -\frac{\mathcal{B}_l}{2}\Big[1 + 2l\Big(\log\frac{2E_0}{\varepsilon} + \psi(l+1)\Big)\Big].
\end{align*}
Here $\psi(z) = \frac{\Gamma'(z)}{\Gamma(z)}$ is the digamma function. Thus, at leading order in $0 < \nu^2 \ll k_1^2$, we find the semiclassical Love numbers,
\begin{align}\label{semicl-Love}
    \boxed{\lambda^{(0)}_{l,\rm sc}
= \left(\frac{G_N}{r_0^2}\right)^{2l+1} \nu^2 \left(2E_0^{-1}E_1\right)^{l}\Big[\frac{\alpha_l}{2}\log \left(2E_0^{-1}E_1\right) + \hat{\lambda}^{\rm ren}_l\left(2E_0^{-1}E_1\right)\Big]} . 
\end{align}

The semiclassical static Love number is a one-loop effect, suppressed by a single power of $l_{\rm pl}^2/r_0^2$
relative to a classical response and multiplied by the classical, dimensionless quantity, $(E_1/M_0)^l\sim(r_0T)^{2l}$. This single power of $\hbar$ arises as $\hbar^{2l+1}$ from $(G_N/r_0^2)^{2l+1}$
compensated by $\hbar^{-2l}$ from $(E_1/E_0)^{l}$. Although our amplitude comes from a tree-level diagram in the near-horizon effective theory, the Schwarzian mode is itself the quantum fluctuation of the ${\rm AdS}_2$ throat, and from the four-dimensional point of view the result is a one-loop quantum-gravitational correction.

Very close to extremality, $k_1 \ll \nu$, the response is analytic,
\begin{align}
    Q(k_1) + 2P(k_1^2)\log\frac{E_0}{\varepsilon}-\frac{
2\pi a\,P(k_1^2)\sin(2\pi\mu)
}{
\cosh(2\pi k_1)-\cos(2\pi\mu)
}
= b_l
+ k_1^{2}b'_l
+\mathcal{O}\!\left(k_1^{4}\right).
\end{align}
Again, absorbing the $c_l$ factor into these coefficients $(c_l a_l, c_l b_l \cdots) \equiv (a_l, b_, \cdots)$, we get:
\begin{align*}
&a_l = 2c_l\pi^2\nu^2 D_l, \quad b_l = a_l\,\hat{S}^{(0)}_l, \quad a'_l = 4T_l\,a_l, \quad
b'_l = a_l\big[4T_l\hat{S}^{(0)}_l + \hat{S}^{(1)}_l\big], \\
&D_l = \prod_{j=1}^{l}\big(\nu^2-j^2\big)^2, \quad
T_l = \sum_{j=1}^{l}\frac{j^2}{(\nu^2-j^2)^2}, \quad
\hat{S}^{(0)}_l = \psi(l+1+\nu) + \psi(\nu-l) + \log\frac{E_0}{\varepsilon}, \\
&\hat{S}^{(1)}_l = -\frac{\psi'(l+1+\nu)+\psi'(\nu-l)}{2\nu} - \frac{\psi''(l+1+\nu)+\psi''(\nu-l)}{2}.
\end{align*}
Thus the leading order static Love numbers close to extremality are given as,
\begin{align}\label{ext-Love}
    \boxed{\lambda^{(0)}_{l,\rm ext}
= \left(\frac{G_N}{r_0^2}\right)^{2l+1}\Big[b_l + \hat{\lambda}^{\rm ren}_l\left(2E_0^{-1}E_1\right)\Big]} . 
\end{align}
As we can see that the energy dependence of this number at extremality is completely scheme dependent by our choice of $\hat{\lambda}^{\rm ren}_l$. A natural choice of this coefficient would be to set it such that the extremal Love number vanishes identically. For finite energy black holes, the Love number has $\log(E_1)$ dependence which cannot vanish for arbitrary black holes by a choice of renormalisation. This is the most important result of this paper.

\section{Discussions} \label{sec: discussions}

In this paper, we have studied the ringdown and tidal deformation effects of near-extremal black holes which are related to scattering processes on the black hole backgrounds. We have incorporated the contributions of universal quantum gravitational fluctuations stemming from the near-horizon geometry of these black holes. We consider a non-rotating extremal background with horizon size $r_0$ and consider near-extremal solutions as deviation of this geometry which are parameterised by the inverse temperature $\beta \gg r_0$. To address the scattering problem on this background, we used the one-dimensional effective description of the quantum gravitational fluctuations localised at the boundary of the near-horizon $AdS_2$ factor. The effective theory consists of Schwarzian and $SU(2)$ modes with a coupling constant $E_0$. We coupled this is effective quantum mechanical theory to an external massless scalar field theory. The coupling is such that a particular mode in the far-horizon quantum field theory acts as a source from the near-horizon perspective and generates a source-operator coupled term in the effective Hamiltonian. The energy $E$ of the Schwarzian states parametrise the mass above extremality and the angular momentum corresponds to rotation above extremality. For the purpose of this paper, we have focused on a near-extremal background with spherical symmetry and kept its energy conserved. 

In this model of the near-extremal black holes, we first computed the leading decay rate of a general Schwarzian+$SU(2)$ state using first-order time-dependent perturbation theory driven by the coupling to the external QFT. We studied the decay in a semiclassical near-extremal regime where $E\gg E_0, \beta \gg r_0$, where it is governed by thermal emission as expected from semiclassical gravity. However, in a quantum near-extremal regime where $E\lesssim E_0, \beta \gg r_0$, the decay is no longer thermal. We find that in both cases, the decay is very small.  

Next, we have considered a $2\to 2$ scattering cross process where the asymptotic states comprise of a state with definite energy and angular momentum in the Schwarzian+$SU(2)$ system and a one-particle state in the external QFT. This is a second-order process in time-dependent perturbation theory. Due to the slow decay of the states, we work in an approximation where the end-states are defined at time scales much larger compared to the interaction time but smaller compared to the decay scale of the external states. The scattering process involved internal states over which we have to integrate. The energies of these internal states get shifted by the self-energy correction which has a small imaginary part. Since our effective theory breaks down at scales $\beta\sim r_0$, we naturally cut off the loop integrals at high energies and find that the real shift in the internal energy is suppressed and it can be neglected for the purposes of this paper. 

Using this corrected internal state energy, we computed the poles of the integrand appearing in the scattering amplitude. We find two different types of poles in the complex internal energy integrals: ones which are independent of the frequency of the external particle and ones that do depend on the frequency. We find the conditions where these two sets of poles can collide on a deformed contour and these conditions give us the poles of the integrated scattering amplitudes. These conditions show that the amplitude diverges at complex frequencies and give us the quasinormal modes of our near-extremal black hole \eqref{QNM}. We find that in the semiclassical limit, the real part is much suppressed compared to the imaginary part and the later matches existing results on zero-damped modes \cite{Hod:2008se, Zimmerman:2015trm}. Interestingly, in the quantum regime, the real part becomes comparable with the imaginary part and may dominate at sufficiently low energies.   

Next, we have computed the conservative response of the $2\to 2$ scattering amplitude and matched it with the tidal-part of the scattering amplitude, associated with static Love numbers in the worldline-EFT to fix the static Love numbers which characterize leading tidal deformability. The amplitude involves an integral over internal states of the Schwarzian theory, which shows UV divergence. To tame the divergences, we analytically continue the conformal dimension of the Schwarzian operators to the complex plane. In a region $0 < {\rm Re}\Delta < 1/2$, the integral converges and gives an analytic expression of the amplitude. We analytically continue these amplitudes near the physical $\Delta = l+1$, where $l$ is the angular momentum of the external scalar. We find that at $l=0$, the static Love number vanishes for any value of energy. This is a consistent with the shift symmetry of the massless scalar. We also find, when the angular momentum threshold is neglected i.e., in the classical limit where the black hole spectrum is smoothly connected to extremality, all static Love numbers vanish \eqref{Love-cl}. This is an important match with the predictions of General Relativity.

Next, we consider the effects of the threshold energy and find nontrivial Love coefficients which show divergences. We show that these divergences can be removed by adding local counterterms on the near-horizon $AdS_2$ boundary involving two insertions of the scalar field. The resultant amplitude is finite, along with an undetermined renormalized static Love number. This is simply reflective of the fact that the Schwarzian theory is to be thought of as a near-horizon EFT with UV cutoff scale determined by $\beta \sim r_0$. The general result for this is shown in \eqref{all-Love}. From this, we analyse the static Love numbers in semiclassical ($E_1 \gg E_0$) and quantum ($E_1 \ll E_0$) regimes. We find that in the semiclassical regime, the Love numbers show a logarithmic dependence on black hole energy, which is a characteristic of the true running of Love numbers. While, very close to extremality, the leading behaviour becomes temperature-independent and does not have a logarithmic dependence on the black hole energy. In fact, in this regime ($k_1 \ll \nu$), absorption is forbidden \cite{Emparan:2025qqf}, the response is purely conservative. However, a natural choice is to fix the scheme-dependent piece $\hat{\lambda}_l^{\rm ren}$ to set the static Love numbers at extremality to zero, similar to the classical case, where the vanishing is governed by $SL(2,R)$ near-horizon symmetry \cite{Charalambous:2022rre}. Although we do not have a concrete justification, but it seems natural because in the quantum theory, the near-horizon $SL(2,R)$ symmetry enhances to an infinite dimensional symmetry that gets explicitly broken by a small temperature. This may be tied to a vanishing of extremal static Love numbers even at the quantum level. However, a complete understanding is beyond the scope of this current paper as for energies exponentially smaller than $E_0$ the Schwarzian description also breaks down. This choice fixes the operator at leading order and shows that as energy increases, the Love numbers are definitely nonzero. Finally, as energy increases but staying in the near-extremal limit, the logarithmic dependence on energy signals a running of the static Love numbers. Similar arguments have been used to show non-zero Love numbers in worldline-EFT frameworks as well~\cite{Saketh:2023bul, Elkhidir:2026lgp}, although there it is usually due to divergences associated with propagation of waves in the curved metric, leading to logarithmic dependence on the perturbation frequency. Here, we observe instead running that depends logarithmically on the energy of the black hole, which can only be fixed by some UV-finite description of quantum black holes. Thus, we have demonstrated that near-horizon quantum gravitational fluctuations will generically induce static Love numbers for non-rotating near-extermal black holes.

We conclude this paper by noting that the quantum gravitational corrections considered in this paper are only localised in the large near-horizon throat of the near-extremal black holes. Typically, in the point particle approximation in WEFT, these modes are not considered. Here quantum corrections can come from far horizon loops governed by the parameter $G_N/r_0^2$ \cite{Elkhidir:2026lgp}, since from very far away, the black hole has only one scale $r_0$. But these effects are not expected to generate temperature-dependent quantum corrections. Unlike Schwarzschild or other non-extremal black holes, the near-horizon throat for a (near-)extremal black hole is very large. Given that this throat supports large quantum gravitational fluctuations, the point particle approximation is not complete in this regime. With our current understanding, it is not obvious how to incorporate both types of quantum corrections in a single unified framework and draw a proper comparison between their contributions. Nevertheless, this is a very important question we hope to answer in future. A similar effective model also should govern the near-horizon dynamics of near-extremal Kerr black holes and their tidal deformations. We would like to compute the QNMs and Love numbers for these black holes as well. We hope to report on this soon.

\acknowledgments
We thank Ashoke Sen for many helpful discussions and his inputs on this work. We also thank P. Ajith, Nabamita Banerjee, Alok Laddha, R. Loganayagam, Godwin Martin, Arnab Rudra and Farman Ullah for useful discussions in various stages of the work. We acknowledge Anthropic's Claude and OpenAI's ChatGPT for helping us evaluate integrals and helping typeset the manuscript. The work of AB is funded by the National Agency for Research and Development ANID - through FONDECYT grant N$^{\circ}$ 3260610. The work of MS and MVSS is supported by Department of Atomic Energy, Government of India, under project no. RTI4019. The research of M.V.S.S. is supported by the National Post-Doctoral Fellowship (PDF/2025/004764), ANRF, Government of India. AB would like to thank Ashoke Sen and ICTS for their hospitality during the initial stages of this work. Finally, we express our gratitude to the people of India for their support in fundamental research.

\appendix

\newpage
\section{Useful identities}\label{App: identities}
\noindent We will use the properties of Gamma function,
\begin{enumerate}
    \item \begin{align}\label{gamma-prod-l}
    \left| \Gamma\!\left(1+l+bi\right) \right|^2
=
\frac{\pi b}{\sinh(\pi b)}
\prod_{j=1}^{l}
\left(j^2+b^2\right), \quad l \in \mathbb{Z}_+
\end{align}

\item \begin{align}\label{gamma-reflection}
    \Gamma(z)\Gamma(1-z) = \frac{\pi}{\sin(\pi z)}.
\end{align}

\item \begin{align}\label{Gamma-large-y}
    |\Gamma(x+iy)| \sim \sqrt{2\pi}|y|^{x-(1/2)}e^{-\pi|y|/2}, \quad \abs{y} \to 
    \infty.
\end{align}
\end{enumerate}

Using the first property, we can rewrite the Schwarzian vertex factor \eqref{Sch-vertex},
\begin{align}\label{Sch-vertex-simple}
    \Gamma^\Delta_{12} =& 2\pi^2\frac{2E_0^{-1}(E_2 - E_1)}{\cosh\left(2\pi\sqrt{2E_0^{-1}E_2}\right)- \cosh\left(2\pi\sqrt{2E_0^{-1}E_1}\right)} \nonumber \\
    \times & \prod_{j=1}^{l} \left[j^2 + \left(\sqrt{2E_0^{-1}E_2} + \sqrt{2E_0^{-1}E_1} \right)^2 \right] \left[j^2 + \left(\sqrt{2E_0^{-1}E_2} - \sqrt{2E_0^{-1}E_1} \right)^2 \right]
\end{align}
We will use the definition of Zeta function,
\begin{align}\label{zeta}
    \zeta(s)
=
\frac{1}{\Gamma(s)}
\int_{0}^{\infty}
\frac{x^{s-1}}{e^x-1}\,dx,
\qquad
\operatorname{Re}(s)>1.
\end{align}

The Principal Value property,
\begin{align}\label{PV}
    \lim_{\epsilon\to 0}\frac{1}{x + i\epsilon} = \mathcal{P}\left(\frac{1}{x}\right) - i\pi \delta(x).
\end{align}

\subsection*{Properties of Wigner 3j-symbols:}
\begin{enumerate}
\item \begin{align}\label{3j-M-sum}
    (2j_3+1)
\sum_{m_1,m_2}
\begin{pmatrix}
j_1 & j_2 & j_3\\
m_1 & m_2 & m_3
\end{pmatrix}
\begin{pmatrix}
j_1 & j_2 & j_3'\\
m_1 & m_2 & m_3'
\end{pmatrix}
=
\delta_{j_3,j_3'}\,
\delta_{m_3,m_3'}.
\end{align}
here, the angular momenta satisfy selection rules.
\item \begin{align}\label{3j-0-column}
    \begin{pmatrix}
L & l_2 & 0\\
M & m_2 & 0
\end{pmatrix}
=
\frac{(-1)^{\,-l_2-m_2}}{\sqrt{1+2l_2}}\,
\delta_{L,l_2}\,
\delta_{M,-m_2}.
\end{align}
\item  The 3j-symbols are invariant under even permutations of the columns, whereas they pick up a factor of $(-1)^{j_1+j_2+j_3}$ under an odd permutation of the columns.
\end{enumerate}

\subsection*{Properties of Wilson function:}

\begin{enumerate}
    \item The series representation of the hypergeometric function is,
\begin{align}
    &{}_pF_q\!\left(
\begin{matrix}
a_1,\ldots,a_p \\
b_1,\ldots,b_q
\end{matrix}
; z
\right)
=
\sum_{m=0}^{\infty}
\frac{
(a_1)_m \cdots (a_p)_m
}{
(b_1)_m \cdots (b_q)_m
}
\frac{z^m}{m!}\\
&(a)_m = \frac{\Gamma(a+m)}{\Gamma(a)} = a(a+1)\cdots(a+m-1)
\end{align}
It converges for Re$(z)<1$.

\item The hypergeometric function appearing in the Wilson function of \eqref{Sch-6j} is of the type $p = q+1, q=3$ and at $z = 1$. This function has poles when the \textit{excess} $s \equiv \sum_{i=1}^q b_i - \sum_{j=1}^p a_j$ has the property that Re$(s)$ is a non-positive integer. The series for $z=1$ converges if Re$(s)>0$.
\end{enumerate}

\section{Derivation of the interaction Hamiltonian}\label{App: Hamiltoninan}
Here, we will derive the interaction \eqref{int-hamil} between the Schwarzian+$SU(2)$ system and the external QFT. Let us begin with the canonically normalised uncharged massless scalar field in four dimensions. The action is given by,
\begin{align}
    S_4 = \frac{1}{2}\int d^4x \sqrt{-g_4} \Phi \Box_4 \Phi.
\end{align}
Here $\Box_4$ denotes the Laplacian on the 4D near-extremal background $g_{4}$,
\begin{align}
    ds^2 = -f(r)dt^2 + f(r)^{-1}dr^2 + r^2 d\Omega^2, \quad f(r) = 1- \frac{2G_N M}{r} + \frac{G_N Q^2}{r^2}.
\end{align}
For our background, $\sqrt{G_N}Q = r_0$ denotes the size of the extremal horizon radius and $G_N M = r_0 + \mathcal{O}(r_0/\beta)$. Using the time translation and spherical symmetry of the background, the classical solution of the scalar can be obtained by reducing the equation of motion to a one-dimensional Schr\"{o}dinger problem in the radial direction. This is given by,
\begin{align}
    &\Phi (t,r,\Omega) = e^{-i\omega t} u(r)Y_{lm}(\Omega), \\
    & \left(\frac{d^2}{dr_*^2}+ \omega^2- \frac{f}{r^2}\left[l(l+1) + r f'\right]\right) r u(r) = 0.
\end{align}
Here $r_*$ is the Tortoise coordinate defined by $dr_* = dr f^{-1}$. This 1D problem can be solved separately in near-horizon and far-horizon regions. The near-horizon solution with ingoing boundary condition at the horizon takes the form, 
\begin{align}
   u_{\rm NH} = P_l^{ - \frac{i\beta\omega}{2\pi}}
\left(
\frac{(-2r_+ + r_+ + r_-)\beta}{4\pi r_+^2} 
\right). 
\end{align}
The near $AdS_2$ boundary behavior of this solution can be obtained by taking a $r\to\infty$ asymptotic limit, where this solution behaves as,
\begin{align}\label{intermediate}
    u_{\rm NH} \xrightarrow[]{r\to\infty} b_1 r^l + b_2 r^{-l-1}.
\end{align}
Here the $b_1$-mode is the non-normalizable piece while the $b_2$-mode is the normalizable one. These coefficients are completely fixed in terms of the parameters of the solution $u_{\rm NH}$,
\begin{align}
    b_1 = \frac{(-1)^{- \frac{i\beta\omega}{4\pi}}2^l\Gamma(l+1/2)}{\sqrt{\pi}\Gamma\left(l+1+\frac{i\beta\omega}{2\pi} \right)} \left(-\frac{\beta}{2\pi r_+^2}\right)^l, \, b_2 = \frac{(-1)^{- \frac{i\beta\omega}{4\pi}}2^{l+1}\Gamma(-2l-1)}{\Gamma(-l)\Gamma\left(-l+\frac{i\beta\omega}{2\pi} \right)} \left(-\frac{\beta}{2\pi r_+^2}\right)^{-l-1}.
\end{align}
For a near-extremal black hole, we can readily see that $b_2/b_1 \to 0$. This corresponds to a decoupling between the near and far-horizon regions.

In the far-horizon region, where we approximate the spacetime as flat, the solution becomes,
\begin{align}
    u_{\rm asymp} = A_{\omega lm} j_l(r\omega) + B_{\omega lm} y_l(r\omega).
\end{align}
The near $AdS_2$ boundary behavior of this solution can be obtained by taking a $r\to 0$ asymptotic limit, where this solution takes a form similar to \eqref{intermediate},
\begin{align}
    u_{\rm asymp} \xrightarrow[]{r\to 0}& \frac{2^{-l-1}\sqrt{\pi}}{\Gamma(l+3/2)}(\omega r)^l A_{\omega lm} - \frac{2^l\Gamma(l+1/2)}{\sqrt{\pi}} (\omega r)^{-l-1} B_{\omega lm} \nonumber \\
    \equiv & c_{\omega lm} r^l + d_{\omega lm} r^{-l-1}.
\end{align}
Matching this solution with \eqref{intermediate}, we find that $c_{\omega lm} = b_1$ and $d_{\omega lm} = b_2 \to 0$. We will work in this decoupling limit. Owing to this decoupling, $u_{\rm asymp}$ essentially behaves as the source term near the $AdS_2$ boundary. The Bessel solution behaves as plane waves close to asymptotic infinity,
\begin{align}
    u_{\rm asymp} \xrightarrow[]{r\to \infty}& \frac{a_{\omega lm}}{\sqrt{4\pi\omega}}\frac{e^{-i\omega r}}{r} + \frac{b_{\omega lm}}{\sqrt{4\pi\omega}}\frac{e^{i\omega r}}{r}, \\
    & \frac{a_{\omega lm}}{\sqrt{4\pi\omega}} = \frac{i^{l+1}A_{\omega lm} + i^l B_{\omega lm}}{2\omega}, \, \frac{b_{\omega lm}}{\sqrt{4\pi\omega}} = \frac{(-i)^{l+1}A_{\omega lm} + (-i)^l B_{\omega lm}}{2\omega} .
\end{align}
In the decoupling limit, $d_{\omega lm}\propto B_{\omega lm} \to 0$ and we get $a_{\omega lm} = (-1)^{l+1} b_{\omega lm}$. This condition equates the free Hamiltonian of the QFT in the past and future null infinities. The classical matching conditions allow us to relate the source mode of the QFT near $AdS_2$ boundary with the plane wave mode of asymptotic infinity through the following relation,
\begin{align}
    c_{\omega lm} = \frac{(2i)^{-l-1}}{\Gamma(l+3/2)}\omega^{l+1/2} a_{\omega lm}.
\end{align}
We define our quantum field theory with the modes $a_{\omega lm}$ corresponding to these plane wave solutions, this is what appears in \eqref{int-hamil}. This matching condition gives us the function $u_{\omega lm}$ appearing in the interaction Hamiltonian. Now we want to find the exact source contribution.

To obtain a canonically normalised scalar in two-dimensions by considering the background to be extremal near-horizon geometry, we can expand the 4D scalar as,
\begin{align}
    \Phi = \frac{1}{r_0}\sum\phi_{lm}(t,r)Y_{lm}(\Omega).
\end{align}
Here $(t,r)$ denote the 2D coordinates on $AdS_2$ throat of size $r_0$ and $Y_{lm}(\Omega)$ are the spherical harmonics on unit $S^2$. The lower-dimensional action takes the form,
\begin{align}
    S_2 = \frac{1}{2}\sum_{lm}\int d^2x \sqrt{-g_2} \left[\phi_{lm} \Box_2 \phi_{lm} - \frac{l(l+1)}{r_0^2}\phi_{lm}^2\right].
\end{align}
This is canonically normalised in 2D. From the 2D throat perspective, the source and the dual operator can be defined from the near-$AdS_2$ boundary behavior of the classical solution in Poincar\'{e} coordinates:
\begin{align}
    \phi_{lm}(t,z) \xrightarrow[]{z\to 0} \left[z^{1-\Delta}J + z^\Delta \mathcal{O}_\Delta\right]e^{-i\omega t} + {\rm h.c.}
\end{align}
Here, $z = \frac{r_0^2}{r}$ is the relation between the Poincar\'{e} coordiante and the usual radial coordinate. $\Delta = l+1$ is the conformal dimension obtained from the effective mass $m^2 = l(l+1)$ and $\omega$ is the dual variable of time. $J,\mathcal{O}_{\Delta lm}$ are the source and dual operator respectively. In terms of the solution in radial coordinate, we can easily identify:
\begin{align}
    J &= \left[r^{-l}r_0^{2l}\phi^{\rm asymp}_{lm}(t,r)\right]_{r\to 0}, \quad \mathcal{O}_{\Delta} = \left[r^{l+1}r_0^{-2l-2}\phi^{\rm asymp}_{lm}(t,r)\right]_{r\to 0}.
\end{align}
Here $\phi^{\rm asymp}$ denotes the solution in the asymptotic region i.e. the Bessel function solution. From dimensional analysis, we can easily check that the combination $J\mathcal{O}_{\Delta lm}$ has the dimensions of energy. The dual operator $\mathcal{O}_{\Delta lm}$ is an operator in the Schwarzian theory and its two-point function must scale in a certain way under conformal transformations. This allows us to rewrite, by abuse of notation, $\mathcal{O}_{\Delta lm} \equiv \quad E_0^\Delta\mathcal{O}_{\Delta lm}$, where the new operator is dimensionless. This redefinition gives us the correct scaling of the two-point function given in \cite{Brown:2024ajk}. Furthermore, we can now expand the asymptotic field in terms of the modes near null infinity and thereby write the source in terms of asymptotic creation and annihilation operators. Now, we write the interaction term in the Hamiltonian as,
\begin{align}
    H_{I} (t) = \sum_{lm}J (t) \mathcal{O}_{\Delta lm} (t) = \sum_{lm} r_0^{2l+1}E_0^\Delta \mathcal{O}_{\Delta lm} (t) \int_0^\infty d\omega \phi_{\omega lm} (t)
\end{align}
In the last step, we have written the mode expansion of the asymptotic field. The prefactor $\mathcal{N}_l = r_0^{2l+1}E_0^\Delta$ appears in \eqref{int-hamil}, where we have further taken out the time dependence of the operator through time evolution factors.

\bibliography{RNlove}
\end{document}